\documentclass[11pt]{article}
\pdfoutput=1
\usepackage[utf8]{inputenc}
\usepackage{dsfont}
\usepackage{IEEEtrantools}
\usepackage{amsfonts}
\usepackage{amsmath}
\allowdisplaybreaks[4]        
\usepackage{amssymb}
\usepackage{upgreek}
\usepackage{euscript}     
\usepackage{braket}
\usepackage{starfont}
\usepackage{color,soul}         
\usepackage{tensor}        
\usepackage{amsthm}
\usepackage{graphicx}
\usepackage{IEEEtrantools}
\usepackage{slashed}
\usepackage{leftidx}
\usepackage{subfigure}
\usepackage{bbm}
\usepackage{bbold}
\definecolor{outerspace}{rgb}{0.25, 0.29, 0.3}
\definecolor{scarlet}{rgb}{1.0, 0.13, 0.0}
\usepackage[header,title,page,titletoc]{appendix}  
\definecolor{princetonorange}{rgb}{1.0, 0.56, 0.0}
\definecolor{WildStrawberry}{rgb}{1.0, 0.26, 0.64}
\definecolor{rossocorsa}{rgb}{0.83, 0.0, 0.0}
\definecolor{navyblue}{rgb}{0.0, 0.0, 0.5}
\usepackage[numbers,sort&compress]{natbib}  
\usepackage{float}
\usepackage[paper=letterpaper,margin=1in]{geometry}
\usepackage{mathtools}
\DeclarePairedDelimiter{\ceil}{\lceil}{\rceil}

\newcommand{\RR}{\mathcal{R}_2}
\newcommand{\RRR}{\mathcal{R}_3}
\newcommand{\Ls}{L_\star}
\newcommand{\bF}[1]{\bar{\mathcal{F}}_{#1}}
\newcommand{\lpeff}{\ell_{ \ssc \rm P}^{\text{eff}}}

\newcommand{\lnr}{^{\mathrm{L}}}

\DeclareMathAlphabet{\pazocal}{OMS}{zplm}{m}{n}
\newcommand{\req}[1]{(\ref{#1})} 

\newcommand{\bea}{\begin{eqnarray}}
\newcommand{\diff}{\mathrm{d}}
\newcommand{\eea}{\end{eqnarray}}
\newcommand{\ba}{\begin{eqnarray}}
\newcommand{\ea}{\end{eqnarray}}

\newcommand{\be}{\begin{equation}}
\newcommand{\ee}{\end{equation} }
\newcommand{\beqa}{\begin{eqnarray}}
\newcommand{\eeqa}{\end{eqnarray}}
\newcommand{\beqar}{\begin{eqnarray*}}
\newcommand{\eeqar}{\end{eqnarray*}}

\renewcommand{\req}[1]{eq.~(\ref{#1})}

\newcommand{\ssc}{\scriptscriptstyle}
\newcommand{\eg}{{\it e.g.,}\ }
\newcommand{\ie}{{\it i.e.,}\ }

\newcommand{\mt}[1]{\textrm{\tiny #1}}

\newcommand{\lp}{\ell_{\mt P}}

\DeclareMathOperator{\tr}{tr}

\makeatletter
\newcommand{\dal}{\mathop{\mathpalette\dal@\relax}}
\newcommand{\dal@}[2]{%
  \begingroup
  \sbox\z@{$\m@th#1\square$}%
  \dimen0=\fontdimen8
    \ifx#1\displaystyle\textfont\else
    \ifx#1\textstyle\textfont\else
    \ifx#1\scriptstyle\scriptfont\else
    \scriptscriptfont\fi\fi\fi3
  \makebox[\wd\z@]{%
    \hbox to \ht\z@{%
      \vrule width \dimen0
      \kern-\dimen0
      \vbox to \ht\z@{
        \hrule height \dimen0 width \ht\z@
        \vss
        \hrule height 2\dimen0
      }%
      \kern-2.5\dimen0
      \vrule width 2.5\dimen0
    }%
  }%
  \endgroup
}
\makeatother

\usepackage{hyperref}
\hypersetup{
    colorlinks,
    citecolor=rossocorsa,
    filecolor=navyblue,
    linkcolor=navyblue,
    urlcolor=navyblue
}

\begin{document} 

\begin{titlepage}
\hfill \\
\begin{flushright}
\hfill{\tt CERN-TH-2022-006}
\end{flushright}

\begin{center}

\phantom{ }
\vspace{0.0cm}

{\bf \Large{Aspects of three-dimensional higher-curvature gravities}}
\vskip 0.5cm
Pablo Bueno,${}^{\text{\Zeus}}$ Pablo A. Cano,${}^{\text{\Kronos}}$ Quim Llorens,${}^{\text{\Apollon}}$ Javier Moreno,${}^{\text{\Hades},\text{\Kronos}}$ and Guido van der Velde${}^{\text{\Admetos}}$
\vskip 0.05in
\small{${}^{\text{\Zeus}}$ \textit{CERN, Theoretical Physics Department,}}
\vskip -.4cm
\small{\textit{CH-1211 Geneva 23, Switzerland.}}

\small{${}^{\text{\Kronos}}$\textit{Instituut voor Theoretische Fysica, KU Leuven}}
\vskip -.4cm
\small{\textit{Celestijnenlaan 200D, B-3001 Leuven, Belgium.}}

${}^{\text{\Apollon}}$ \textit{Departament de F\'isica Qu\`antica i Astrof\'isica, Institut de Ci\`encies del Cosmos,}\\ \textit{Universitat de
Barcelona, Mart\'i i Franqu\`es 1, E-08028 Barcelona, Spain}

\small{${}^{\text{\Hades}}$ \textit{Instituto de F\'isica, Pontificia Universidad Cat\'olica de Valpara\'iso}}
\vskip -.4cm
\small{\textit{Casilla 4059, Valpara\'iso, Chile.}}

\small{${}^{\text{\Admetos}}$ \textit{Instituto Balseiro, Centro At\'omico Bariloche}}
\vskip -.4cm
\small{\textit{ 8400-S.C. de Bariloche, R\'io Negro, Argentina.}}

\begin{abstract}
We present new results involving general higher-curvature gravities  in three dimensions. The most general Lagrangian of that kind can be written as a function of $R,\mathcal{S}_2,\mathcal{S}_3$, where $R$ is the Ricci scalar, $\mathcal{S}_2\equiv \tilde R_{a}^b \tilde R_b^a$, $\mathcal{S}_3\equiv \tilde R_a^b \tilde R_b^c \tilde R_c^a$, and $\tilde R_{ab}$ is the traceless part of the Ricci tensor. First, we provide a general formula for the exact number of independent order-$n$ densities, $\#(n)$. This satisfies the identity $\#(n-6)=\#(n)-n$. Then, we show that, linearized around a general Einstein solution, a generic order-$n\geq 2$ density can be written as a linear combination of $R^n$, which by itself would not propagate the generic massive graviton, plus a density which by itself would not propagate the generic scalar mode, $ R^n-12n(n-1)R^{n-2}\mathcal{S}_2$, plus $\#(n)-2$ densities which contribute trivially to the linearized equations. Next, we obtain an analytic formula for the quasinormal modes and frequencies of the BTZ black hole as a function of the masses of the graviton and scalar modes for a general theory. Then, we provide a recursive formula as well as a general closed expression for order-$n$ densities which non-trivially satisfy an holographic c-theorem, clarify their relation with Born-Infeld gravities and prove that the scalar mode is always absent from their spectrum. We show that, at each order $n \geq 6$, there exist $\#(n-6)$ densities which satisfy the holographic c-theorem in a trivial way and that all of them are proportional to a single sextic density $\Omega_{(6)}\equiv 6 \mathcal{S}_3^2-\mathcal{S}_2^3$. Next, we show that there are also $\#(n-6)$ order-$n$ Generalized Quasi-topological densities in three dimensions, all of which are ``trivial'' in the sense of making no contribution to the metric function equation. Remarkably, the set of such densities turns out to coincide exactly with the one of theories trivially satisfying the holographic c-theorem. We comment on the meaning of $\Omega_{(6)}$ and its relation to the Segre classification of three-dimensional metrics.  




\end{abstract}
\end{center}
\end{titlepage}

\setcounter{tocdepth}{2}

{\parskip = .2\baselineskip \tableofcontents}


\section{Introduction}
\label{sec:Introduction}
Gravity becomes simpler when we move from four to three dimensions. Firstly, the Weyl tensor vanishes identically, implying that all curvatures are Ricci curvatures. This means that all solutions of three-dimensional Einstein gravity are locally equivalent to maximally symmetric backgrounds and that no gravitational waves propagate. In spite of this, global differences between spacetimes  do appear and prevent the theory from being ``trivial'', even at the classical level. In particular, in the presence of a cosmological constant, the theory admits black hole solutions \cite{Banados:1992wn,Banados:1992gq} which, despite important differences with their higher-dimensional counterparts, do share many of their properties ---including the existence of event and Cauchy horizons, thermodynamic properties, holographic interpretation, etc. 

The local equivalence of all classical solutions allows for a characterization of the phase space of the theory \cite{Witten:1988hc}.  In addition ---up to non-negligible details--- three-dimensional Einstein gravity is classically equivalent to a Chern-Simons gauge theory \cite{Achucarro:1986uwr}. From an holographic point of view \cite{Maldacena,Witten,Gubser}, these qualitative changes with respect to higher dimensions are manifest in the distinct nature of conformal field theories in two dimensions. In fact, while the observation that the symmetry algebra  of AdS$_3$ spaces is generated  by two copies of the conformal algebra in two dimensions  \cite{Brown:1986nw}   is often considered to be  a precursor of AdS/CFT, the nature of the putative holographic theory ---or ensemble of theories--- dual to pure Einstein gravity is still subject of debate \cite{Maloney:2007ud,Keller:2014xba,Benjamin:2019stq,Alday:2019vdr,Cotler:2020ugk,Maxfield:2020ale}.

The above simplifications also affect higher-curvature modifications of Einstein gravity. In particular, all theories can be constructed exclusively from contractions of the Ricci tensor, which reduces the number of independent densities drastically. Similarly, the usual arguments for considering higher-curvature corrections ---which involve their appearance in the form of infinite towers of terms coming from stringy corrections--- do not make much sense in three-dimensions. This is because all non-Riemann curvatures can be removed via field redefinitions, and hence one is left again with Einstein gravity ---plus cosmological constant and a possible gravitational Chern-Simons term \cite{Gupta:2007th}. However, there is a different reason to consider higher-curvature gravities with non-perturbative couplings in three dimensions. This is the fact that, as opposed to Einstein gravity, they can give rise to non-trivial local dynamics. This  appears in the form of a massive graviton and/or a scalar mode ---see \eg \cite{Gullu:2010sd}.


 By far, the best known higher-curvature modification of Einstein gravity in three dimensions is the so-called ``New Massive Gravity'' (NMG) \cite{Bergshoeff:2009hq}.\footnote{Interestingly, the NMG quadratic density constructed in \cite{Bergshoeff:2009hq} had been identified in the mathematical literature \cite{Gursky} years before the seminal paper appeared. We thank Bayram Tekin for pointing this out to us. } At the linearized level, the theory describes a massive graviton  with the same dynamics of a Fierz-Pauli theory. In addition, the theory is distinguished by possessing second-order traced equations \cite{Oliva:2010zd}, by admitting an holographic c-theorem \cite{Sinha:2010ai} and by admitting a Chern-Simons description \cite{Hohm:2012vh}. Unfortunately, demanding unitarity of the bulk theory spoils the unitarity of the boundary theory and viceversa \cite{Bergshoeff:2009aq}, a problem which has been argued to be unavoidable for general higher-curvature theories sharing the spectrum of NMG \cite{Gullu:2010vw}.
 
Moving from quadratic to higher orders, one can use some of the above criteria to select special theories. One possibility is to demand that the corresponding theories admit an holographic c-theorem \cite{Sinha:2010ai,Paulos:2010ke}. Alternatively, one can look for additional theories which admit a Chern-Simons description \cite{Afshar:2014ffa,Bergshoeff:2014bia,Bergshoeff:2021tbz}. A different route involves considering special $D\rightarrow 3$ limits of higher-dimensional theories with special properties \cite{Alkac:2020zhg}. Often, the densities resulting from these different approaches coincide with each other.   Alternative routes include \cite{Banados:2009it,Paulos:2012xe,Bergshoeff:2013xma,Bergshoeff:2014pca,Alkac:2017vgg,Alkac:2018eck,Ozkan:2018cxj,Afshar:2019npk}. 
 
While higher-curvature modifications of three-dimensional Einstein gravity have been studied extensively by now, most of the results are only valid for the lowest curvature orders or for particular theories ---see \eg \cite{Paulos:2010ke,Gurses:2011fv,Gurses:2015zia} for exceptions. In this paper we present a collection of new results for general-order higher-curvature theories. Without further ado, let us summarize them.




\subsection{Summary of results}\label{summary}
\begin{itemize}
\item In Section \ref{counting}, we obtain a formula for the exact number of independent order-$n$ densities, $\#(n)$. This is given by
  \begin{equation}\label{num2}
  \#(n) = \ceil[\Big]{\frac{n}{2} \left(\frac{n}{6}+1 \right)+\epsilon}\, ,
  \end{equation}
  where $\ceil[]{x}$ is the usual ceiling function and $\epsilon$ is any positive number such that $\epsilon\ll 1$. The function $\#(n)$ satisfies the interesting recursive relation $\#(n-6)=\#(n)-n$, which says that the number of order-$n$ densities minus $n$ equals the number of densities of six orders less.
  \item In Section \ref{eomEs} we present the equations of motion for a general higher-curvature gravity and the algebraic equations these reduce to when evaluated for Einstein metrics. We also make a few comments about single-vacuum theories.
  \item In Section \ref{seclineq} we obtain the linearized equations of a general higher-curvature gravity around an Einstein spacetime as a function of the effective Planck length and the masses of the new spin-2 and spin-0 modes generically propagated. Formulas for such physical parameters are obtained for a general theory. Using these results, we show that the most general order-$n$ density can be written as 
  \begin{equation}
  \mathcal{L}_{(n)}=\alpha_n R^n+ \beta_n [R^n-12 n (n-1) R^{n-2}\mathcal{S}_2]+\mathcal{G}_{(n)}^{\rm trivial}\, ,
  \end{equation}
  where: the first term a density which by itself does not propagate the massive spin-2 mode but which does propagate the scalar one (for $n\geq 2$), the second term is a density which by itself does not propagate the scalar mode but which does propagate the spin-2 mode, and the third term, $\mathcal{G}_{(n)}^{\rm trivial}$ ---which involves $\#(n)-2$ densities--- does not contribute at all to the linearized equations.
  \item In Section \ref{BTZsec} we use the results of the previous section to obtain explicit formulas for the Quasi-normal modes and frequencies of the BTZ black hole in a general higher curvature gravity.
 \item In Section \ref{ctheorem}  we study higher-curvature theories which satisfy an holographic c-theorem. First, we provide a recursive formula for densities which satisfy it in a non-trivial fashion ---\ie they contribute non-trivially to the c-function. This is given by
 \begin{equation}\label{recuu}
\mathcal{C}_{(n)}=\frac{4(n-1)(n-2)}{3n(n-3)}\left(\mathcal{C}_{(n-1)}\mathcal{C}_{(1)}-\mathcal{C}_{(n-2)}\mathcal{C}_{(2)}\right)\, ,
\end{equation}
 which allows one to obtain an order-$n$ density of that kind from the two immediately lower order ones. Then, we solve the recurrence explicitly and provide a general explicit formula for  $\mathcal{C}_{(n)}$. We argue that there are $\#(n)-(n-1)$ densities of order $n$ which satisfy the holographic c-theorem. Of those, $\#(n)-n$ are trivial in the sense of making no contribution to the c-function. We show that all such ``trivial'' densities are proportional to the sextic density
 \begin{equation}
 \Omega_{(6)}\equiv 6\mathcal{S}_3^2-\mathcal{S}_2^3\, ,
 \end{equation}
 so that the most general order-$n$ density satisfying the holographic c-theorem can be written as
 \begin{equation}\label{cThLs}
\mathcal{L}_{(n)}^{\rm c-theorem}=\alpha_n \mathcal{C}_{(n)}+\Omega_{(6)}\cdot \mathcal{L}^{\rm general}_{(n-6)}\, ,
\end{equation}
where $ \mathcal{L}^{\rm general}_{(n-6)}$ is the most general linear combination of order-$(n-6)$ densities. In addition, we show that if a theory satisfies the holographic c-theorem, then it does not include the scalar mode in its spectrum. Finally, we study the relation of $\mathcal{C}_{(n)}$ with the general term obtained from expanding the Born-Infeld gravity Lagrangian of \cite{Gullu:2010pc}.
\item In Section \ref{GQTss} we explore the possible existence of Generalized Quasi-topological gravities in three dimensions. We show that there exist $\#(n)-n$ theories of that kind, and that all of them are ``trivial'' ---in the sense of making no contribution to the equation of the black hole metric function--- and again proportional to the same sextic density $ \Omega_{(6)}$ that appeared in the previous section.
\item In Section \ref{Ommm6} we make some comments on the relation between $\Omega_{(6)}$ and the Segre classification of three-dimensional spacetimes. We explain why the prominent role played by this density in the identification of ``trivial'' densities of the types studied in the previous two sections could have been expected ---at least to some extent.
\end{itemize}

\subsection{Notation and conventions}
Throughout the paper we consider higher-curvature theories constructed from contractions of the Ricci tensor and the metric. We use latin indices $a,b,c,\dots$ for world tensors. When referring to generic Lagrangian densities, we use the notation $\mathcal{L}\equiv \mathcal{L}(g_{ab},R_{ab})$, and we express the gravitational constant in terms of the Planck length $\ell_{\rm \ssc P}\equiv 8\pi G_{\rm N}$.  We choose to work always with a negative cosmological constant which we denote in terms of the action length scale $L$, so that  $-2\Lambda\equiv 2/L^2$. The Anti-de Sitter$_3$ (AdS$_3$) radius is denoted by $L_{\star}$, and sometimes we use the notation $\chi_{0}\equiv L^2/L_{\star}^2$, so that $\chi_{0}=1$ for Einstein gravity ---in the notation of \eg \cite{Buchel:2009sk,Myers:2010jv,Bueno:2018xqc}, $\chi_{0}\equiv f_{\infty}$. We will often consider Lagrangians which involve an Einstein gravity plus cosmological constant part, plus a general function of the three basic densities which span the most general higher-curvature invariants in three-dimensions. Those three invariants can be alternatively chosen to be $\{R,\mathcal{R}_2\equiv R_{a}^b R_{b}^a, \mathcal{R}_3\equiv R_a^bR_b^cR_c^a\}$ or $\{R,\mathcal{S}_2\equiv \tilde R_{a}^b \tilde R_{b}^a, \mathcal{S}_3\equiv \tilde R_a^b \tilde R_b^c \tilde R_c^a\}$, where $\tilde R_{ab}$ is the traceless part of the Ricci tensor. Consequently, we will often consider general functions of either set of densities, which we will denote respectively by $\mathcal{F}\equiv \mathcal{F}(R,\mathcal{R}_2,\mathcal{R}_3)$ and $\mathcal{G}\equiv \mathcal{G}(R,\mathcal{S}_2,\mathcal{S}_3)$. The different invariants are often classified attending to their curvature order $n$, corresponding to the number of Ricci tensors involved in their definition. Generic order-$n$ densities are denoted $\mathcal{L}_{(n)}$  and the most general linear combination of order-$n$ densities is denoted $\mathcal{L}_{(n)}^{\rm general}$. We use the notation $\mathcal{G}_X\equiv \partial \mathcal{G}/\partial X$, $\mathcal{G}_{X,X}\equiv \partial^2 \mathcal{G}/\partial X^2$ and so on to denote partial derivatives. Expressions with a bar denote evaluation of the invariants on an Einstein background metric,  $\bar X \equiv X(\bar R, \bar{\mathcal{R}}_2,\bar{\mathcal{R}}_3)$. In the case of terms  which require taking derivatives with respect to some of the arguments, it is understood that the derivatives are taken first, and the resulting expression is then evaluated on the background. Hence, for instance, $\bar{\mathcal{F}}_R \equiv \left. [\partial \mathcal{F}/\partial R]\right|_{R= \bar R,\mathcal{R}_2= \bar{\mathcal{R}}_2,\mathcal{R}_3=\bar{\mathcal{R}}_3}$.

\section{Counting higher-curvature densities}\label{counting}

In this section, we compute the exact number of independent densities of order $n$ constructed from arbitrary contractions of the Riemann tensor and the metric.  The vanishing of the Weyl tensor in three dimensions reduces the analysis to theories constructed from contractions of the Ricci tensor and the metric,  $\mathcal{L}\left(g_{ab},R_{ab}\right)$. Additionally, the existence of Schouten identities  implies that the most general higher-curvature action can be written as \cite{Paulos:2010ke,Gurses:2011fv}
\begin{equation}\label{action3}
S_{(\mathcal{R})}=\frac{1}{2\ell_{\rm \ssc P}} \int \diff ^3x \sqrt{|g|} \mathcal{L}_{(\mathcal{R})}\, , \quad \mathcal{L}_{(\mathcal{R})}\equiv \frac{2}{L^2}+R+\mathcal{F}\left(R,\mathcal{R}_2,\mathcal{R}_3 \right) \, ,
\end{equation}
where we chose a negative cosmological constant and we defined 
\begin{equation}
\mathcal{R}_2\equiv R_a^bR_b^a\, , \quad \mathcal{R}_3\equiv R_a^bR_b^c R_c^a\, .
\end{equation}
%
Often we will assume $\mathcal{F}\left(R,\mathcal{R}_2,\mathcal{R}_3 \right) $ to be either an analytic function of its arguments, or a series of the form
\begin{equation}\label{Fseries}
\mathcal{F}\left(R,\mathcal{R}_2,\mathcal{R}_3 \right) =\sum_{i,j,k}L^{2(i+2j+3k-1)}\,  \alpha_{ijk}  R^i \mathcal{R}_2^j \mathcal{R}_3^k\, ,
\end{equation}
for some dimensionless coefficients $\alpha_{ijk}$.

As we just mentioned, the ``Schouten identities'' drastically reduce the number of independent densities of a given order, leaving \req{action3} as the most general case. Those identities take the form \cite{Paulos:2010ke}
\begin{equation}\label{schou}
\delta_{b_1 \dots b_n}^{a_1\dots a_n} R_{a_1}^{b_1}R_{a_2}^{b_2}\cdots R_{a_n}^{b_n}=0\, ,\quad \text{for} \quad n>3\, ,
\end{equation}
and rely on the fact that totally antisymmetric tensors with ranks higher than $3$ vanish identically in $D=3$. From \req{schou}, it follows that the cyclic contraction of $n>3$ Riccis can be written in terms of lower-order densities, and hence the generality of \req{action3}. One finds, for instance
%
%
\begin{align}
R_a^bR_b^cR_c^dR_d^a &=  \frac{1}{6} R^4+ \frac{4}{3} R \mathcal{R}_3 + \frac{1}{2} \mathcal{R}_2^2- \mathcal{R}_2 R^2 \, , \\  R_a^bR_b^cR_c^dR_d^e R_e^a&= \frac{1}{6}R^5+\frac{5}{6}\left(\mathcal{R}_3\mathcal{R}_2+\mathcal{R}_3 R^2-\mathcal{R}_2 R^3\right)\, , \\
R_a^bR_b^cR_c^dR_d^e R_e^fR_f^a&= \frac{1}{12}R^6+\mathcal{R}_3\mathcal{R}_2 R+\frac{1}{3}\mathcal{R}_3 R^3-\frac{1}{4}\mathcal{R}_2 R^4 -\frac{3}{4}\mathcal{R}_2^2 R^2+\frac{1}{4}\mathcal{R}_2^3+ \frac{1}{3} \mathcal{R}_3^2\, . 
\end{align}

It is often convenient to use a basis of invariants involving the traceless part of the Ricci tensor,
\begin{equation}\label{tracelessR}
\tilde R_{ab}\equiv R_{ab}-\frac{1}{3}g_{ab} R\, .
\end{equation}
Then, we can define 
\begin{equation}
\mathcal{S}_2\equiv \tilde R_a^b \tilde R_b^a=\mathcal{R}_2-\frac{1}{3}R^2\, , \quad  \mathcal{S}_3\equiv \tilde R_a^b \tilde R_b^c \tilde R_c^a=\mathcal{R}_3-R \mathcal{R}_2+\frac{2}{9}R^3\, ,
\end{equation}
and alternatively write the most general theory replacing $\mathcal{F}\left(R,\mathcal{R}_2,\mathcal{R}_3 \right) $ by $\mathcal{G}\left(R,\mathcal{S}_2,\mathcal{S}_3 \right)$ in \req{action3} \cite{Gurses:2011fv}, namely
\begin{equation}\label{action4}
S_{(\mathcal{S})}=\frac{1}{2\ell_{\rm \ssc P}} \int \diff ^3x \sqrt{|g|} \mathcal{L}_{(\mathcal{S})}\, , \quad \mathcal{L}_{(\mathcal{S})}\equiv \frac{2}{L^2}+R+\mathcal{G}\left(R,\mathcal{S}_2,\mathcal{S}_3 \right) \, .
\end{equation}
We will write the polynomial version of $\mathcal{G}$ as
\begin{equation}\label{Gseries}
\mathcal{G}(R,\mathcal{S}_2,\mathcal{S}_3)=\sum_{i,j,k}L^{2(i+2j+3k-1)}\,  \beta_{ijk}  R^i \mathcal{S}_2^j \mathcal{S}_3^k\, .
\end{equation}
While \req{action3}  feels like a  more natural choice from a higher-dimensional perspective, it turns out that many formulas simplify considerably when expressed in terms of $\tilde R_{ab}$ instead. We will try to present most of our results in both bases.

Let us consider the case in which the theory is a power series of the building blocks $R,\mathcal{R}_2,\mathcal{R}_3$ (or, alternatively, $R,\mathcal{S}_2,\mathcal{S}_3$). The order $n$ of a certain combination of scalar invariants is related to the powers of the individual components through $n=i+2j+3k$. One finds the following possible invariants at the first orders,
\begin{align}
R\, , \quad \text{for}\quad &n=1\, ,\\
R^2\, , \quad \mathcal{R}_2\, , \quad \text{for}\quad &n=2\, ,\\
R^3\, , \quad R \mathcal{R}_2\, , \quad \mathcal{R}_3\, ,\quad \text{for}\quad &n=3\, ,\\
R^4\, , \quad R^2 \mathcal{R}_2\, , \quad R \mathcal{R}_3\, , \quad \mathcal{R}_2^2 \, , \quad \text{for}\quad &n=4\, ,\\
R^5\, , \quad R^3 \mathcal{R}_2\, , \quad R^2 \mathcal{R}_3\, , \quad R \mathcal{R}_2^2 \, , \quad \mathcal{R}_2 \mathcal{R}_3\, , \quad \text{for}\quad &n=5\, ,\\
R^6\, , \quad R^4 \mathcal{R}_2\, , \quad R^3 \mathcal{R}_3\, , \quad R^2 \mathcal{R}_2^2 \, , \quad R \mathcal{R}_2 \mathcal{R}_3\, , \quad \mathcal{R}_2^3\, ,\quad \mathcal{R}_3^2\, , \quad \text{for}\quad &n=6\, ,
\end{align}
and so on. Then, the function $\#(n)$ counting the number of invariants of order $n$ takes the values $\#(1)=1$, $\#(2)=2$, $\#(3)=3$, $\#(4)=4$, $\#(5)=5$, $\#(6)=7$. 


In order to find the explicit form of $\#(n)$ as a function of $n$, we can proceed as follows. If we understand the number of elements constructed from powers of $R$ alone up to order $n$ as the coefficients of a power series, we can define the generating function $f^{(R)}(x)$ as 
\begin{equation}
f^{(R)}(x)\equiv \frac{1}{1-x}\sim 1+x+x^2+x^3+\dots\, ,
\end{equation}
\ie such that the rhs, which is the Maclaurin series of the lhs, has coefficient $1$ for all powers. This is because at every order $n$ there is a single density we can construct with $R$ alone, namely, $R^n$. Now, if we want to do the same for $\mathcal{R}_2$, we need to take into account that the corresponding coefficients should be $1$ when $n$ is even, and $0$ otherwise. We define then
\begin{equation}
f^{(\mathcal{R}_2)}(x)\equiv  \frac{1}{1-x^2}\sim 1+x^2+x^4+x^6+\dots\, 
\end{equation}
Following the same reasoning for $\mathcal{R}_3$, we define
\begin{equation}\label{eq:GenFR}
f^{(\mathcal{R}_3)}(x)\equiv  \frac{1}{1-x^3}\sim 1+x^3+x^6+x^9+\dots\, 
\end{equation}  
 Now, we can obtain $\#(n)$ as the coefficient of the Maclaurin series corresponding to the generating function which results from the product of the three generating functions previously defined, namely
 \begin{equation}
 f^{(R)}(x) f^{(\mathcal{R}_2)}(x) f^{(\mathcal{R}_3)}(x)=\frac{1}{(1-x)(1-x^2)(1-x^3)}\sim \sum_n \#(n) x^n \, .
 \end{equation}
The result can be written explicitly as
\begin{equation}\label{num1}
\#(n)=\frac{1}{72}\left[47+(-1)^n9+6n(6+n)+16\cos\left(\frac{2n\pi}{3}\right)\right]\, .
\end{equation}
This gives the exact number of independent  three-dimensional order $n$ densities. It is easy to verify that this yields the same values obtained above for the first $n$'s. Note that $\#(n)$ is not an analytic function, but it is still easy to see that it goes as
 \begin{equation}\label{apro}
 \#(n)\sim \frac{n}{2} \left(\frac{n}{6}+1 \right)\, ,
  \end{equation}
  for $n\gg 1$. The fact that $ \#(n)$ scales with $\sim n^2$ for large $n$ had been previously observed in \cite{Paulos:2010ke}.
  
  It can be shown that $\#(n)$ can be alternatively written exactly (for integer $n$, which is the relevant case) as
  \begin{equation}\label{num2}
  \#(n) = \ceil[\Big]{\frac{n}{2} \left(\frac{n}{6}+1 \right)+\epsilon}\, ,
  \end{equation}
  where $\ceil[]{x}\equiv {\rm min} \left\{k \in \mathbb{Z}\, |\, k\geq x \right\}$ is the usual ceiling function and $\epsilon$ is any positive number such that $\epsilon \ll 1$. For instance, at order $n=1729$, one has $\#(1729)=249985$ independent densities, as one can easily verify both from \req{num1} or \req{num2}. 
  
 The function $\#(n)$ satisfies several relations which connect its values at different orders. A particularly suggestive one is the recursive relation
 \begin{align}\label{properttt}
   \#(n-6) &=  \#(n)-n\, ,
 \end{align}
 which connects the number of densities of a given order with the number of densities  of six orders less. This follows straightforwardly from the general expression of $\#(n)$ in \req{num2}. We will use this relation in Sections \ref{ctheorem} and \ref{GQTss} to prove a couple of results concerning the general form of densities which trivially satisfy an holographic c-theorem and of densities which belong to the Generalized Quasi-topological class.



  

\section{Equations of motion and Einstein solutions}\label{eomEs}
The equations of motion of a general higher-curvature theory constructed from arbitrary contractions of the Ricci scalar and the metric can be written as \cite{Padmanabhan:2011ex}
\begin{equation}\label{eomsss}
\mathcal{E}_{ab}\equiv P_{a}^{c}R_{b c} -\frac{1}{2}g_{a b}\mathcal{L}-\nabla_{(a}\nabla_c P_{b)}^c+\frac{1}{2}\dal P_{a b}+\frac{1}{2}g_{a b} \nabla_c\nabla_d P^{c d}=0\, , \quad \text{where} \quad P^{a b}\equiv \left.\frac{\partial \mathcal{L}}{\partial R_{a b}}\right\vert_{g^{c d}}\, .
\end{equation}
In our three-dimensional case, when written as in \req{action3}, the explicit form of these equations reads 
\begin{equation}\label{eq:EOM2}
\begin{aligned}
\mathcal{E}_{ab}^{(\mathcal{R})}\equiv &+R_{a b}(1+\mathcal{F}_R)-\frac{1}{2}g_{a b}\left(R+\frac{2}{L^2}+\mathcal{F}\right)+\left(g_{a b}\dal - \nabla_{a}\nabla_{b}\right) \mathcal{F}_R\\ &+ 2 \mathcal{F}_{\mathcal{R}_2}  R_{a}^{c}R_{c b}+ 3 \mathcal{F}_{\mathcal{R}_3}R_{a}^{c}R_{c d}R_{b}^{d}+g_{a b}\nabla_{c}\nabla_{d}\left(\mathcal{F}_{\mathcal{R}_2}R^{c d}+\frac{3}{2}\mathcal{F}_{\mathcal{R}_3}R^{c f}R_{f}^{d}\right)\\
&+\dal \left( \mathcal{F}_{\mathcal{R}_2}R_{a b}+\frac{3}{2}\mathcal{F}_{\mathcal{R}_3}R_{a}^{c}R_{c b}\right)-2\nabla_{c}\nabla_{(a}\left(R_{b)}^{c}\mathcal{F}_{\mathcal{R}_2}+\frac{3}{2}R_{b)}^{d}R_{d}^{c}\mathcal{F}_{\mathcal{R}_3}\right)=0 \, ,
\end{aligned}
\end{equation}
In the $R,\mathcal{S}_2,\mathcal{S}_3$ basis, the equations of motion read instead \cite{Gurses:2011fv}
\begin{equation}\label{eq:EOM3}
\begin{aligned}
\mathcal{E}_{ab}^{(\mathcal{S})}\equiv &+\left(\tilde{R}_{a b}+\frac{1}{3}g_{a b}R\right)-\frac{1}{2}g_{a b}\left(R+\frac{2}{L^2}+\mathcal{G}\right)+2 \mathcal{G}_{\mathcal{S}_2}  \tilde{R}_{a}^{c}\tilde{R}_{c b}+ 3 \mathcal{G}_{\mathcal{S}_3}\tilde{R}_{a}^{c}\tilde{R}_{c d}\tilde{R}_{b}^{d}\\ &+\left(g_{a b}\dal - \nabla_{a}\nabla_{b}+\tilde{R}_{a b}+\frac{1}{3}g_{a b}R\right) \left(\mathcal{G}_R-\mathcal{G}_{\mathcal{S}_3}\mathcal{S}_2\right)+g_{a b}\nabla_{c}\nabla_{d}\left(\mathcal{G}_{\mathcal{S}_2}S^{c d}+\frac{3}{2}\mathcal{G}_{\mathcal{S}_3}S^{c f}\tilde{R}_{f}^{d}\right)\\
&+\left(\dal+\frac{2}{3}R\right) \left( \mathcal{G}_{\mathcal{S}_2}\tilde{R}_{a b}+\frac{3}{2}\mathcal{G}_{\mathcal{S}_3}\tilde{R}_{a}^{c}\tilde{R}_{c b}\right)-2\nabla_{c}\nabla_{(a}\left(\tilde{R}_{b)}^{c}\mathcal{G}_{\mathcal{S}_2}+\frac{3}{2}\tilde{R}_{b)}^{d}\tilde{R}_{d}^{c}\mathcal{G}_{\mathcal{S}_3}\right)=0 \, .
\end{aligned}
\end{equation}

Solutions of Einstein gravity plus cosmological constant can be easily embedded in the general higher-curvature theory 
\req{action3} or \req{action4}. These include, for instance, pure AdS$_3$ and the BTZ black hole. Indeed, consider Einstein metrics of the form 
\begin{equation}\label{constantr}
\bar R_{ab}=-\frac{2}{L_{\star}^2} \bar g_{ab}\, .
\end{equation}
In that case, one has
\begin{equation}\label{RR2R3S2S3}
\bar R=-\frac{6}{L_{\star}^2}\, , \quad \bar{\mathcal{R}}_2=\frac{12}{L_{\star}^ 4}\, , \quad   \bar{\mathcal{R}}_3=-\frac{24}{L_{\star}^ 6}\, , \quad \bar{\mathcal{S}}_2=0\, , \quad \bar{\mathcal{S}}_3=0\, .
\end{equation}
Hence, \req{constantr} satisfies the equations of motion \req{eq:EOM2} provided
\begin{equation} \label{sks}
\frac{6}{ L^2}-\frac{6}{L_{\star}^2}\left[1-2\bar{\mathcal{F}}_R+\frac{8}{L_{\star}^2 } \bar{\mathcal{F}}_{\mathcal{R}_2}-\frac{2 4}{L_{\star}^4 } \bar{\mathcal{F}}_{\mathcal{R}_3} \right] +3\bar{\mathcal{F}}=0
\, ,
\end{equation}
is satisfied. In the alternative formulation in terms of traceless Ricci tensors, the analogous equation is considerably simpler and reads \cite{Gurses:2011fv}\begin{equation}\label{vacuS}
\frac{6}{ L^2}-\frac{6}{L_{\star}^2}\left[1- 2 \bar{\mathcal{G}}_R \right] +3\bar{\mathcal{G}}=0\,.
\end{equation}
For Einstein gravity, this simply reduces to $L^2=L_{\star}^2$, which just says that the AdS$_3$ radius coincides with the cosmological constant scale. In general, \req{sks} and \req{vacuS} are equations for the quotient $\chi_{0}\equiv L^2/L_{\star}^2$. 
If the series form (\ref{Fseries}) is assumed, \req{sks} takes the form 
\begin{equation}
1-\chi_{0}+\sum_{n} a_n  \chi_{0}^n=0\, ,\quad \text{where}\quad a_n\equiv (-1)^n 6^{n-1} (3-2n) \sum_{j,k} \frac{\alpha_{n-2j-3k,j,k} }{3^{j+2k}} \, .
\end{equation}
Similarly, \req{vacuS} takes the form
\begin{equation}
1-\chi_{0}+\sum_{n} b_n  \chi_{0}^n=0\, ,\quad \text{where}\quad b_n\equiv (-1)^n 6^{n-1} (3-2n) \beta_{n00} \,,
\end{equation}
where observe that terms involving $\mathcal{S}_2$ and $\mathcal{S}_3$ make no contribution to the equation.


On general grounds, the above polynomial equations will possibly have several positive solutions for $\chi_{0}$, so the corresponding theories will possess several AdS$_3$ vacua.  Finding higher-curvature theories with a single vacuum in three and higher dimensions has been subject of study of numerous papers ---see \eg \cite{Crisostomo:2000bb,Gullu:2015cha,Karasu:2016ifk} and references therein. In the present case, a complete analysis of the conditions which lead to a single vacuum can be easily performed in a case-by-case basis, but not so much for a completely general theory, so we will not pursue it here.  Let us nonetheless make a couple of comments. First, observe that all extensions of Einstein gravity with terms involving either $\mathcal{S}_2$ and/or $\mathcal{S}_3$ will have a single vacuum, since for those the Einstein gravity solution $\chi_{0}=1$ will be the only one.  A different possibility for single-vacuum theories would correspond to an order-$n$ degeneration of the solutions of the above polynomial equations, \ie to the cases in which these become
\begin{equation}
\left(1-\frac{\chi_0}{n}\right)^n=0\, .
\end{equation} 
Observe that this  involves $n-1$ conditions for a theory containing densities of order $n$ and lower and these will necessarily mix couplings of different orders. In particular, for a theory written in the $\{R,\mathcal{S}_2,\mathcal{S}_3 \}$ basis involving densities of order up to $n$, these read
\begin{equation}
\beta_{i00}=\binom{n}{i}\frac{1}{n^i 6^{i-1}(3-2i)}\, , \quad i=2,\dots, n\,.
\end{equation}
Hence, a Lagrangian of the form
\begin{align}
\mathcal{L}_{(n)}^{\rm single\, \, vac.}&=\frac{2}{L^2}+R+\sum_{i=2}^n \binom{n}{i}\frac{L^{2(i-1)}}{n^i 6^{i-1}(3-2i)}R^i + \mathcal{S}_2 h_2(R,\mathcal{S}_2,\mathcal{S}_3)+ \mathcal{S}_3 h_3(R,\mathcal{S}_2,\mathcal{S}_3)\, ,
\end{align}
where $h_{2,3}$ are any analytic functions of their arguments will have a single AdS$_3$ vacuum.


\section{Linearized equations}\label{seclineq}
The linearized equations of motion around maximally symmetric backgrounds of higher-curvature gravities involving general contractions of the Riemann tensor and the metric  were obtained in \cite{PabloPablo,Aspects} ---see also \cite{Tekin1,Tekin2}. The resulting expression was expressed in terms of four parameters, $a$, $b$, $c$ and $e$, and a simple method for computing such coefficients for a given theory was also provided, along with the connection between them and the relevant physical parameters ---namely, the effective Newton constant and the masses of the additional modes. In this section we apply this method to a general higher-curvature theory in three dimensions and classify theories according to the content of their linearized spectrum. 

Let $g_{ab}=\bar g_{ab}+h_{ab}$ where the background metric is an Einstein spacetime satisfying \req{constantr} and 
$h_{ab}\ll 1$, $\forall a,b=0,1,2$. Then, restricted to a general three-dimensional higher-curvature gravity of the form \req{action3}, the equations of motion of the theory read, at leading order in the perturbation
 \cite{Aspects}  
 \begin{equation}\label{line}
\frac{1}{4\ell_{\ssc \rm P}}\mathcal{E}\lnr_{ab}\equiv \left[e+c\left(\bar{\dal}+\frac{2}{L_{\star}^2}\right)\right]G\lnr_{ab}+(2b+c)\left(\bar{g}_{ab}\bar{\dal}-\bar{\nabla}_a\bar{\nabla}_b\right)R\lnr-\frac{1}{L_{\star}^2}\left(4b+c\right)\bar{g}_{ab}R\lnr=\frac{1}{4}T_{ab}\lnr\, ,
\end{equation}
where we included a putative matter stress-tensor for clarity purposes and where the linearized Einstein and Ricci tensors, and Ricci scalar read
 \begin{IEEEeqnarray}{ll}
G\lnr_{ab}&=R\lnr_{ab}-\frac{1}{2}\bar{g}_{ab}R\lnr+\frac{2}{L_{\star}^2} h_{ab}\, ,\\
R\lnr_{ab}&=\bar{\nabla}_{\left(a\right|}\bar{\nabla}_{c}h\indices{^c_{\left|b\right)}}-\frac{1}{2}\bar{\dal}h_{ab}-\frac{1}{2}\bar{\nabla}_a\bar{\nabla}_b h-\frac{3}{L_{\star}^2} h_{ab}+\frac{1}{L_{\star}^2} h \bar{g}_{ab}\, ,\\
R\lnr &=\bar{\nabla}^a\bar{\nabla}^b h_{ab}-\bar{\dal} h+\frac{2}{L_{\star}^2} h\, .
\end{IEEEeqnarray}
 In higher dimensions there is an additional parameter ---denoted ``$a$'' in \cite{Aspects}--- appearing in the linearized equations. However, this turns out to be nonzero only for densities which involve Riemann curvatures, and so we have $a=0$ for all three-dimensional theories. For a generic higher-curvature theory in that case, \req{line} describes three propagating degrees of freedom corresponding to a massive ghost-like spin-2 mode plus a spin-0 mode.  The parameters $e$, $c$ and $b$ above can be related to the effective Planck length $\ell_{\ssc \rm P}^{\rm eff}$ and the masses (squared) of such modes, which we denote $m_g^2$ and $m_s^2$, as
\begin{equation}\label{phypara}
\ell_{\ssc \rm P}^{\rm eff}=\frac{1}{4e}\, , \quad m_g^2=-\frac{e}{c}\, , \quad m_s^2=\frac{e+\frac{8}{L_{\star}^2}(3b+c)}{3c+8b}\, .
\end{equation}
In subsection \ref{classifi} we explain how to compute these parameters for a general higher-curvature theory and do this explicitly in our three-dimensional context.

In terms of the physical quantities, the linearized equations read
\begin{equation}
\begin{aligned}\label{linEQ}
\frac{\ell_{\ssc \rm P}^{\rm eff}}{\ell_{\ssc \rm P}} m_g^2\cdot \mathcal{E}\lnr_{ab}\equiv &+ \left(m_g^2-\frac{2}{L_{\star}^2}-\bar{\dal}\right)G\lnr_{ab}- \frac{1}{L_{\star}^2}\left(\frac{m_g^2+m_s^2-\frac{2}{L_{\star}^2}}{2(m_s^2-\frac{3}{L_{\star}^2})} \right) \bar g_{ab} R\lnr \\  &+ \left(\frac{m_g^2-m_s^2+\frac{4}{L_{\star}^2}}{4(m_s^2-\frac{3}{L_{\star}^2})} \right) (\bar g_{ab} \bar \dal- \bar\nabla_a\bar\nabla_b)R\lnr=\ell_{\ssc \rm P}^{\rm eff}m_g^2\cdot T\lnr_{ab}\, .
\end{aligned}
\end{equation}

\subsection{Physical modes}\label{physmod}
From what we have said so far, it is not obvious that \req{linEQ} describes the aforementioned modes of masses $m_s$, $m_g$. In order to see this, it is convenient to decompose the metric perturbation as 
\begin{equation}
h_{ab}=\hat{h}_{ab}+\frac{\bar\nabla_{\langle a}\bar\nabla_{b\rangle} h}{\left(m_s^2-\frac{3}{L_{\star}^2}\right)}+\frac{1}{3}h \bar{g}_{ab}\, ,
\end{equation}
where $\langle ab\rangle$ denotes the traceless part, and $\hat{h}_{ab}$ satisfies 
\begin{equation}
\bar{g}^{ab}\hat{h}_{ab}=0\, ,\quad \bar\nabla^{a}\hat h_{ab}=0\, ,
\end{equation}
where the second condition is imposed using gauge freedom. Let us note that this decomposition fails in the special case $m_s^2=\frac{3}{L_{\star}^2}$. In that situation, it is not possible to decouple the trace and traceless parts of $h_{ab}$. However, from \req{phypara}, it follows that in this case $m_g^2=-1/L_\star^2$, so the spin-2 mode is a tachyon. Hence, we will assume that $m_s^2\neq\frac{3}{L_{\star}^2}$ to avoid this problematic situation. 

Then, the trace and the traceless part of the linearized equations become, respectively, \cite{Aspects}
\begin{align}\label{hmode}
\frac{2}{L_{\star}^2} \frac{\left(1+\frac{1}{m_g^2L_{\star}^2}\right)}{\left(m_s^2-\frac{3}{L_{\star}^2}\right)} (\bar \dal - m_s^2)h &= \ell_{\ssc \rm P}^{\rm eff}T^{\rm L}\, , \\ \label{hmode2}
\frac{1}{2m_g^2} \left(\bar \dal +\frac{2}{L_{\star}^2} \right)\left(\bar \dal +\frac{2}{L_{\star}^2}-m_g^2\right) \hat{h}_{ab}&= \ell_{\ssc \rm P}^{\rm eff} T^{\rm{L,eff}}_{\langle ab\rangle}\, ,
\end{align}
where $T^L\equiv \bar g^{ab}T_{ab}^L$ and 
\begin{equation}
T^{\rm{L,eff}}_{\langle ab\rangle}\equiv T^{\rm{L}}_{\langle ab\rangle}-\frac{L_{\star}^2}{2}\frac{\left(\bar \dal +\frac{1}{L_{\star}^2}-m_g^2 \right)}{\left(m_g^2+\frac{1}{L_{\star}^2} \right)}\bar\nabla_{\langle a}\bar\nabla_{b\rangle} T^{\rm L}\, .
\end{equation}
Eq. (\ref{hmode}) describes a spin-0 mode corresponding to the trace of the perturbation. On the other hand, \req{hmode2} can be further rewritten by defining
\begin{equation}
\hat h_{ab}\equiv \hat h_{ab}^{(m)} +\hat h_{ab}^{(M)}\, ,\quad \text{where} \quad \hat h_{ab}^{(m)}\equiv -\frac{1}{m_g^2}\left[ \bar \dal +\frac{2}{L_{\star}^2}-m_g^2\right] \hat h_{ab}\, , \quad \hat h_{ab}^{(M)}\equiv \frac{1}{m_g^2}\left[ \bar \dal +\frac{2}{L_{\star}^2}\right] \hat h_{ab}\, , 
\end{equation}
as
\begin{align}
-\left(\bar \dal +\frac{2}{L_{\star}^2}\right)\hat h_{ab}^{(m)} &= \ell_{\ssc \rm P}^{\rm eff} T^{\rm{L,eff}}_{\langle ab\rangle} \, , \\ \label{spin2m}
\left(\bar \dal +\frac{2}{L_{\star}^2}-m_g^2\right)\hat h_{ab}^{(M)} &=  \ell_{\ssc \rm P}^{\rm eff} T^{\rm{L,eff}}_{\langle ab\rangle}\, .
\end{align}
These describe two traceless spin-2 modes which couple to matter with opposite signs. However, as opposed to higher dimensions, only the massive one is propagating in $D=3$. The would-be massless spin-2 mode is pure gauge (whenever $T^{L}_{ab}=0$) in this number of dimensions  ---see \eg \cite{Deser:1983tn,Nakasone:2009bn,Myung:2011bn,Moynihan:2020ejh}.\footnote{Massless and massive gravitons in $D$ dimensions propagate  $\frac{D(D-3)}{2}$ and $\frac{(D+1)(D-2)}{2}$ degrees of freedom, respectively, which means $0$ and $2$ degrees of freedom respectively for $D=3$.}  Hence, the relevant equations are \req{hmode} and \req{spin2m} which describe a maximum of three degrees of freedom ---one from the scalar mode and two from the spin-2 one--- propagated around Einstein solutions by higher-curvature gravities in the most general case. 

When $\ell_{\ssc \rm P}^{\rm eff}>0$, the massive graviton is a ghost and the scalar mode has positive energy, but since there is no massless graviton, one could also consider $\ell_{\ssc \rm P}^{\rm eff}<0$, so that the massive graviton has positive energy and the scalar is a ghost. As we will see below, there are theories that only propagate either the scalar mode or the massive spin-2 mode, and these can be made unitary by taking $\ell_{\ssc \rm P}^{\rm eff}>0$ or $\ell_{\ssc \rm P}^{\rm eff}<0$, respectively. An example of the latter is NMG as introduced in \cite{Bergshoeff:2009hq}, in which the Ricci scalar appears with the ``wrong'' sign, hence implying $\ell_{\ssc \rm P}^{\rm eff}<0$.

\subsection{Identification of physical parameters}
Given a higher-curvature theory, one can linearize its equations and deduce the values of the parameters $b,c,e$ (and consequently $\ell_{\rm \ssc P}^{\rm eff},m_g^2,m_s^2$) by comparing them with the above general expressions. A much faster way of performing this  identification was proposed in \cite{Aspects}, which we adapt here to our three-dimensional setup. One starts by replacing all Ricci tensors in the Lagrangian by
\begin{equation}
R^{\rm aux}_{ab}=-\frac{2}{L_{\star}^2}g_{ab}+\alpha (x -1)k_{ab} \, ,
\end{equation}
where $x$ is an arbitrary integer constant and the symmetric tensor $k_{ab}$ is defined such that $k_a^a\equiv x$ and $k_a^b k_b^c=k_a^c$. Then, the parameters can be unambiguously extracted from the general formulas \cite{Aspects}
\begin{equation}
\left. \frac{\partial \mathcal L(R^{\rm aux}_{ab})}{\partial \alpha}\right|_{\alpha=0}=2e\, x(x-1)\, , \quad \left. \frac{\partial^2 \mathcal L(R^{\rm aux}_{ab})}{\partial \alpha^2}\right|_{\alpha=0}=4x(x-1)^2 (c+ b x)\, .
\end{equation}
It is straightforward to do this for our general three-dimensional actions. When the theory is expressed in terms of the traceless Ricci tensor as in \req{action4}, the resulting parameters take a particularly simple form
\begin{equation}\label{GGL}
e=\frac{1}{4\ell_{\ssc \rm P}} [1+\bar{\mathcal{G}}_R]\, , \quad b=\frac{1}{4\ell_{\ssc \rm P}} \left[\frac{1}{2}\bar{\mathcal{G}}_{R,R}-\frac{1}{3} \bar{\mathcal{G}}_{\mathcal{S}_2} \right]\, , \quad c=\frac{1}{4\ell_{\ssc \rm P}} \bar{\mathcal{G}}_{\mathcal{S}_2}\, ,
\end{equation}
where recall that we are using the notation $\mathcal{G}_X\equiv \partial \mathcal{G}/\partial X$, $\mathcal{G}_{X,X}\equiv \partial^2 \mathcal{G}/\partial X^2$ and the bar means that we are evaluating the resulting expressions on the background geometry, which is implemented through \req{RR2R3S2S3}. 
If we assume that the Lagrangian allows for a polynomial expansion, it is useful to decompose $\mathcal{G}$ in the following way,
\begin{equation}\label{GS2}
\mathcal{G}(R,\mathcal{S}_2,\mathcal{S}_3)=f(R)+\mathcal{S}_2 g(R)+\mathcal{G}_{\rm triv}\, , \quad \text{where} \quad \mathcal{G}_{\rm triv}\equiv \mathcal{S}_2^2 h(R,\mathcal{S}_2)+\mathcal{S}_3 l(R,\mathcal{S}_2,\mathcal{S}_3)
\end{equation}
includes all terms which do not contribute to the linearized equations around any constant curvature solution. That is the case of any density involving any power of $\mathcal{S}_2$ greater or equal than two and any power of $\mathcal{S}_3$ (different from zero).
With the Lagrangian expressed in this way, the background equation, \req{vacuS}, reduces to
\begin{equation}
\frac{6}{L^2}-\frac{6}{L_{\star}^2}[1-2 \bar f_R]+3\bar f=0\, ,
\end{equation}
and using eq. \eqref{phypara} we find the physical quantities of the linearized spectrum, 
\begin{align}\label{ppog}
\ell_{\rm \ssc P}^{\rm eff}=\frac{\ell_{\rm \ssc P}}{[1+\bar f_R]}\, ,\quad
m_g^2=-\frac{[1+\bar f_R]}{ \bar g}\, ,\quad
m_s^2=\frac{[1+\bar f_R]+\frac{12}{L_{\star}^2} \bar f_{R,R}}{4\bar f_{R,R}+\frac{1}{3}\bar g}\, .
\end{align}
When expressed explicitly in terms of the gravitational couplings in an expansion of the form (\ref{Gseries}) these read 
\begin{equation}
\begin{aligned}\label{ppog2}
\ell_{\rm \ssc P}^{\rm eff}=&\frac{\ell_{\rm \ssc P}}{\left[1+\sum_i\beta_{i00} i(-6 \chi_{0})^{i-1} \right]}\, ,\quad
m_g^2=-\frac{\left[ 1+\sum_i\beta_{i00} i(-6\chi_{0})^{i-1} \right]}{\Ls^2\sum_i\beta_{i10}(-6)^i}\\
m_s^2=&\frac{\left[1-\sum_i\beta_{i00} i (2 i-3)(-6\chi_{0})^{n-1} \right]}{4\Ls^2\sum_i(-6 \chi_{0})^{i-2}  [(i-1) i \beta_{i00}+3\beta_{i10}]}\, .
\end{aligned}
\end{equation}

With the above expressions at hand, it is straightforward to classify the different theories according to the presence or absence of the massive graviton and scalar modes in their spectrum. Before doing so, let us present in passing the expressions analogous to \req{ppog} when the analysis is performed for a theory expressed in the 
$\{R,\mathcal{R}_2,\mathcal{R}_3\}$ basis instead. In that case, the equations become more involved and a decomposition of the form (\ref{GS2}) is not available. We have  
\begin{equation}
\begin{aligned}\label{555}
\lpeff=&\frac{\lp\Ls^4}{\Ls^4(1+\bF{R})-4\Ls^2\bF{\RR}+12\bF{\RRR}}\, , \quad
m_g^2=\frac{\Ls^4(1+\bF{R})-4\Ls^2\bF{\RR}+12\bF{\RRR}}{6\Ls^2\bF{\RRR}-\Ls^4\bF{\RR}}\, , \\
m_s^2=&\frac{3}{\Ls^2}+\left[\Ls^8\left(1+\bF{R}\right)-5\Ls^6\bF{\RR}-18\Ls^4\bF{\RRR}\right]/\left[3\Ls^8\bF{\RR}-18\Ls^6\bF{\RRR}+4\Ls^8\bF{R,R}\right.\\
&\left.-32\Ls^6\bF{R,\RR}+96\Ls^4\bF{R,\RRR}-384\Ls^2\bF{\RR,\RRR}+4\Ls^8\bF{R,R}+64\Ls^4\bF{\RR,\RR}+576\bF{\RRR,\RRR}\right]\, .
\end{aligned}
\end{equation}
The polynomial form is straightforward to obtain from these expressions (and as ugly as one may anticipate).

\subsection{Classification of theories}\label{classifi}
The decomposition (\ref{GS2}) and \req{ppog} make it very simple to classify all theories depending on the mode content of their linearized spectrum. The three sets of theories we consider are: theories which are equivalent to Einstein gravity at the linearized level, theories which do not propagate the massive graviton, and theories which do propagate the scalar mode.

\subsubsection{Einstein-like theories}
A first group of densities are those for which $m_g^2,m_s^2\rightarrow \infty$, namely, densities in whose spectrum both the massive graviton and the scalar mode are absent. These are theories which, at the level of the linearized equations, are identical to Einstein gravity ---up to, at most, a change in the effective Planck length.  As we mentioned earlier, a large set of densities do not contribute whatsoever to the linearized equations. These are given by 
\begin{equation}\label{gtriv}
\mathcal{G}|_{\text{trivial linearized equations}}= \mathcal{S}_2^2 h(R,\mathcal{S}_2)+\mathcal{S}_3 l(R,\mathcal{S}_2,\mathcal{S}_3)\, .
\end{equation}
It is not difficult to see that there are $\#(n)-2$ densities of this kind at order $n$. Namely, all order-$n$ densities but those of the forms $R^n$ and $\mathcal{S}_2 R^{n-2}$ contribute trivially to the linearized equations. While there are no ``trivial'' densities for $n=1,2$, they start to proliferate for $n\geq 3$, becoming the vast majority for higher orders. As it turns out, these ``trivial densities'' are the only Einstein-like theories which exist beyond Einstein gravity itself. The reason is that removing both the massive graviton and the scalar from the spectrum amounts at imposing $c=b=0$, which implies $\bar g=\bar f_{R,R}=0$. These are on-shell conditions, but if we want to avoid relations between densities of different orders, we must force them to hold for any value of $\bar R$. Hence, the conditions become $g(R)=f_{R,R}(R)\equiv 0$, whose only non-trivial solution besides \req{gtriv} is Einstein gravity plus a cosmological constant.  Hence, most higher-curvature densities have in fact trivial linearized equations.

It is a remarkable ---and exclusively three-dimensional--- fact that Einstein gravity is unique in this sense. Observe that starting in four dimensions and for higher $D$ there are generally several Einstein-like densities with non-trivial linearized equations at each curvature order. Examples are Lovelock \cite{Lovelock1,Lovelock2} and some $f(\text{Lovelock})$ densities \cite{Love}, Einsteinian cubic gravity \cite{PabloPablo}, Quasi-topological \cite{Quasi2,Quasi,Dehghani:2011vu,Cisterna:2017umf} and Generalized quasi-topological gravities \cite{Hennigar:2017ego,PabloPablo3,PabloPablo4,Bueno:2019ycr}, among others \cite{Li:2017ncu,Karasu:2016ifk,Li:2017txk}.

\subsubsection{Theories without massive graviton}
Theories for which  $m^2_g\rightarrow\infty$ do not propagate the massive graviton. In terms of our parameters $e$, $b$ and $c$, this condition is given by $c=0$.
From \req{ppog} it is clear that this set of theories are those with $\bar g\equiv g(\bar R)=0$. Again, in order to impose this condition at each curvature order we must demand $g(R)\equiv 0$. Hence, the most general (polynomial) density which makes a non-trivial contribution to the linearized equations and which does not propagate the massive graviton in three-dimensions is $f(R)$ gravity
\begin{equation}
\mathcal{G}|_{\text{no massive graviton}}=f(R)\, ,
\end{equation}
Obviously, at  order $n$ there is $1$ such density, corresponding to $R^n$. Of course, one can obtain more complicated densities satisfying the $m^2_g\rightarrow\infty$ condition by combining some of the trivial Einstein-like densities with the $f(R)$ ones. Hence, there are actually $\#(n)-1$ independent densities which do not propagate the massive graviton at order $n$.

For comparison, observe that in $D\geq 4$ there is a large set of higher-curvature theories which do not have the massive graviton in their spectrum. This is the case, in particular, of all $f(\text{Lovelock})$ theories \cite{Love} ---the set also includes all the Einstein-like theories mentioned in the last paragraph of the previous subsubsection. 

\subsubsection{Theories without scalar mode}
The condition for the scalar mode to be absent from the spectrum, $m_s^2\rightarrow \infty$,  reads instead $3c+8b=0$, which is satisfied by theories for which $12\bar f_{R,R}+\bar g=0$. From this we learn that the most general class theories of this kind contributing non-trivially to the linearized equations reads
\begin{equation}
\mathcal{G}|_{\text{no scalar mode}}=f(R)-12f_{R,R}(R)\mathcal{S}_2\, ,\qquad (f_{R,R}(R)\neq 0)
\end{equation}
Again, there is a single order-$n$ density of this kind, corresponding to
\begin{align}
\mathcal{G}^{(n)}|_{\text{no scalar mode}} &=R^n-12 n (n-1) R^{n-2}\mathcal{S}_2 \, , \\  &= [1+4n(n-1)]R^n-12 n (n-1) R^{n-2}\mathcal{R}_2\, .
\end{align}
For $n=2$, the above density is nothing but the New Massive Gravity one \cite{Bergshoeff:2009hq}.  Once again, we can combine the above order-$n$ densities with the $\#(n)-2$ ``trivial'' densities to obtain additional densities which do not propagate the scalar mode. There are then $\#(n)-1$ densities which do not propagate the scalar mode at each order.

In higher dimensions, a prototypical example of a theory which satisfies this condition  is conformal gravity \cite{Hassan:2013pca,Aspects}, which can be thought of as a natural $D$-dimensional extension of NMG. 

In sum, in $D=3$, at any order  $n\geq 2$ we can always decompose the most general linear combination of higher-curvature densities as a sum of a term which by itself would not propagate the massive graviton, plus a term which by itself would not propagate the scalar mode, plus  $\#(n)-2$ densities which do not contribute to the masses of any of them. 

\section{The BTZ black hole in higher-order gravity}\label{BTZsec}
Since the BTZ black hole \cite{Banados:1992wn,Banados:1992gq} is an Einstein spacetime ---and hence it is locally AdS$_3$--- it is an exact solution of all higher-derivative theories of gravity. The parameters of the solution have different values, though.  To begin with, the AdS scale is not given by the cosmological constant scale $L$, but by $L_{\star}=L/\sqrt{\chi_{0}}$, as we discussed earlier. Thus, the general, rotating BTZ metric can be written as
\begin{equation}\label{eq:rotatingBTZ}
\diff s^2=-f(r)\diff t^2+\frac{\diff r^2}{f(r)}+r^2\left(\diff \phi-\frac{r_{+}r_{-}}{L_{\star} r^2}\diff t\right)^2\, , \quad \text{where}\quad
f(r)\equiv \frac{(r^2-r_{+}^2)(r^2-r_{-}^2)}{L_{\star}^2r^2}\, ,
\end{equation}
and we assume that $r_{+}^2>r_{-}^2$. In addition, the coordinate $\phi$ has periodicity $2\pi$. 

In this section we study the quasinormal modes (QNMs) of the BTZ solutions for general higher-curvature theories. In order to do that, we first need to identify the mass and angular momentum of the solution in this general context, and for that we can resort to thermodynamics. 

\subsection{Thermodynamics}
The horizon of the BTZ black hole is located at $r=r_{+}$, and this leads to a Hawking temperature
\begin{equation}
T=\frac{f'(r_{+})}{4\pi}=\frac{r_{+}^2-r_{-}^2}{2\pi L_{\star}^2r_{+}}\, .
\end{equation}
On the other hand, it will also be important to take note of the angular velocity of the horizon,
\begin{equation}
\Omega=\frac{r_{-}}{L_{\star}r_{+}}\, .
\end{equation}
We can find the free energy by evaluating the Euclidean on-shell action, whose bulk part is 
\begin{equation}
I_{E}^{\rm bulk}=-\frac{1}{2\ell_{\rm \ssc P}} \int_{\mathcal{M}} \diff ^3x \sqrt{g} \mathcal{L}\, .
\end{equation}
On the other hand, for asymptotically AdS solutions (as in this case), one can use the prescription given in \cite{Bueno:2018xqc} ---see also \cite{Araya:2021atx}--- for the boundary terms and counterterms,
\begin{equation}
I_{E}^{\rm bdry}=-\frac{a^*}{\pi L_{\star}}\int_{\partial M}\diff^2x\sqrt{h}\left(K-\frac{1}{L_{\star}}\right)\, ,
\quad \text{where} \quad
a^{*}=-\frac{\pi L_{\star}^3}{4\ell_{\rm \ssc P}}\bar{\mathcal{L}}\, ,
\end{equation}
and $K$ is the usual Gibbons-Hawking-York term \cite{York:1972sj,Gibbons:1976ue} corresponding to the trace of the extrinsic curvature of $\partial \mathcal{M}$.
In the holographic context, the constant $a^{*}$ represents the universal contribution to the entanglement entropy across a spherical entangling region \cite{Myers:2010tj,Myers:2010xs,CHM} in general dimensions which, for general higher-curvature gravities, is proportional to the on-shell Lagrangian \cite{Imbimbo:1999bj,Schwimmer:2008yh,Myers:2010tj,Myers:2010xs,Bueno:2018xqc}. In the present case, $a^{*}$ is proportional to the central charge of the two-dimensional dual CFT, $c=12 a^*$. For the Lagrangian expressed as in eq.~\eqref{action4}, we have
\begin{equation}
a^{*}=\frac{\pi L_{\star}^3}{4\ell_{\rm \ssc P}}\left(\frac{6}{L_{\star}^2}-\frac{2}{L^2}-\bar{\mathcal{G}}\right)\, ,
\end{equation}
and after making use of the relation between $L$ and $L_{\star}$ \eqref{vacuS}, we can write it as
\begin{equation}
a^{*}=\frac{\pi L_{\star}}{\ell_{\rm \ssc P}}\left(1+\bar{\mathcal{G}}_{R}\right)=\frac{\pi L_{\star}}{\ell_{\rm \ssc P}^{\rm eff}}\, .
\end{equation}

Therefore, the Euclidean action on the BTZ black hole is simply computed by
\begin{equation}
I_{E}=-\frac{\bar{\mathcal{L}}}{2\ell_{\rm \ssc P}} \left[\int_{\mathcal{M}} \diff ^3x \sqrt{g}-\frac{L_{\star}^2}{2}\int_{\partial M}\diff^2x\sqrt{h}\left(K-\frac{1}{L_{\star}}\right)\right]\, ,
\end{equation}
which gives the following value for the free energy $F\equiv I_{E}/\beta$,
\begin{equation}
F=-\frac{\pi(r_{+}^2-r_{-}^2)}{\ell_{\rm \ssc P}^{\rm eff}L_{\star}^2}\, .
\end{equation}
Now, the entropy can be computed from the Wald's formula \cite{Wald:1993nt,Iyer:1994ys}, which gives
\begin{equation}
S=-\frac{\pi}{\ell_{\rm \ssc P}}\int d\sigma \frac{\partial \mathcal{L}}{\partial R^{ab}}\epsilon_{ac}\tensor{\epsilon}{_{b}^{c}}=\frac{2\pi}{\ell_{\rm \ssc P}}\left(1+\bar{\mathcal{G}}_{R}\right)A\, ,
\end{equation}
where  $A=2\pi r_{+}$ is the area of the horizon. Therefore, this result can be written as
\begin{equation}
S=\frac{4\pi^2 r_{+}}{\ell_{\rm \ssc P}^{\rm eff}}\, .
\end{equation}

Combining the entropy and the free energy, we can obtain the mass, 
\begin{equation}
M=F+TS=\frac{\pi(r_{+}^2+r_{-}^2)}{\ell_{\rm \ssc P}^{\rm eff}L_{\star}^2}\, .
\end{equation}
Finally, from the first law
\begin{equation}
\diff M-T\diff S=\frac{2\pi r_{-}}{\ell_{\rm \ssc P}^{\rm eff}L_{\star}^2 r_{+}}\left(\diff r_{-}r_{+}+\diff r_{+}r_{-}\right)=\Omega \diff\left(\frac{2\pi r_{+}r_{-}}{\ell_{\rm \ssc P}^{\rm eff}L_{\star}}\right)\, ,
\end{equation}
we identify the angular momentum,
\begin{equation}
J=\frac{2\pi r_{+}r_{-}}{\ell_{\rm \ssc P}^{\rm eff}L_{\star}}\, .
\end{equation}
Thus, everything turns out to work out just like for Einstein gravity, with the identifications $\ell_{\rm \ssc P}\rightarrow \ell_{\rm \ssc P}^{\rm eff}$ and $L\rightarrow L_{\star}$.

\subsection{Perturbations and quasinormal modes}
While the BTZ black hole is a solution of all higher-order gravities, its perturbations behave differently depending on the theory. In  fact, for Einstein gravity it makes no sense to consider the gravitational perturbations around a BTZ black hole, since all of them are trivial (equivalent to infinitesimal gauge transformations), due to the fact that all solutions of three-dimensional Einstein gravity are locally AdS$_3$. 
However, as we saw, higher-derivative gravities do introduce additional degrees of freedom: in general a scalar mode and a massive graviton.  Interestingly, since the BTZ is locally AdS$_3$, we can apply our results from section~\ref{seclineq} and perform a general analysis valid for all higher-curvature gravities. 

Following the results obtained in subsection \ref{physmod}, the quasinormal modes are solutions to the equations 
\begin{align}
\left(\bar{\dal}-m_s^2\right)h=0\, ,\quad
\left(\bar{\dal}+\frac{2}{L_{\star}^2}-m_g^2\right)\hat{h}_{ab}=0\, ,
\label{eq:massiveeq}
\end{align}
which satisfy  an outgoing-wave boundary condition at the black hole horizon, and that behave as normalizable modes at infinity. The QNMs of a scalar field in the background of the BTZ black hole are well-known~\cite{Cardoso:2001hn,Birmingham:2001pj,Berti:2009kk}, while those of the massive graviton were computed in \cite{Myung:2011bn}  in the case of NMG. However, in a general metric theory we have both types of perturbations, with masses that depend on the different parameters of the Lagrangian. Let us then compute the quasinormal modes of these perturbations and their associated frequencies.

\subsubsection{The unit radius BTZ}
To simplify matters, we can consider the BTZ black hole with $r_{+}=L_{\star}$, $r_{-}=0$. The general case can be obtained from this one after appropriate coordinate transformations. The metric of this black hole reads
\begin{equation}\label{eq:unitBTZ}
\diff s^2=-\left(\frac{r^2}{L_{\star}^2}-1\right)\diff t^2+\frac{\diff r^2}{\left(\frac{r^2}{L_{\star}^2}-1\right)}+r^2\diff \phi^2\, ,
\end{equation}
and introducing 
$
r=L_{\star}\cosh(\rho)\, ,
$
it can be written as
\begin{equation}\label{eq:unitBTZ2}
\diff s^2=-\sinh^2(\rho) \diff t^2+L_{\star}^2\cosh^2(\rho) \diff \phi^2+L_{\star}^2 \diff \rho^2\, .
\end{equation}
Finally, one can also write this metric in terms of the null coordinates $u=t+L_{\star}\phi$, $v=t-L_{\star}\phi$, so that we have
\begin{equation}
\diff s^2=\frac{1}{4}\left(\diff u^2+\diff v^2\right)-\frac{1}{2}\cosh(2\rho)\diff u \diff v+L_{\star}^2\diff \rho^2\, .
\end{equation}
The interesting property about these coordinates is that the two sets of generators of the $\mathrm{SL}(2,\mathbb{R})$ algebra, $\{L_{k}\}$ and $\{\bar L_{k}\}$, with $k=-1,0,1$, take a symmetric form, namely,
\begin{align}
L_0&=-\partial_{u}\, ,\quad L_{\pm 1}=e^{\pm u}\left(-\frac{\cosh(2\rho)}{\sinh(2\rho)}\partial_{u}-\frac{1}{\sinh(2\rho)}\partial_{v}\pm\frac{1}{2 L_{\star}}\partial_{\rho}\right)\, ,\\
\bar L_0&=-\partial_{v}\, ,\quad \bar L_{\pm 1}=e^{\pm v}\left(-\frac{\cosh(2\rho)}{\sinh(2\rho)}\partial_{v}-\frac{1}{\sinh(2\rho)}\partial_{u}\pm\frac{1}{2 L_{\star}}\partial_{\rho}\right)\, .
\end{align}
Then, the idea of reference \cite{Sachs:2008gt} is that the QNMs of the BTZ black hole can be found as the descendants of ``chiral highest weight" modes, which are those that are either annihilated by $L_{1}$ or by $\bar L_{1}$. These are also the fundamental QNMs, \textit{i.e.}, those with the lowest (in magnitude) imaginary part. 

In the case of a scalar field (the trace of the metric $h$ in our case), we separate the dependence of the field on the $u$ and $v$ variables and consider a perturbation of the form
\begin{equation}
h=e^{-i up_{+}-iv p_{-}} F(\rho)\, .
\end{equation}
Then we first search for the chiral highest weight modes, $h_{(0)}^{L}$ and $h_{(0)}^{R}$, which satisfy 
\begin{equation}
L_{1}h_{(0)}^{L}=0\, ,\qquad \bar{L}_{1}h_{(0)}^{R}=0\, ,
\end{equation}
corresponding to left-moving and right-moving modes, respectively. These are first order equations for the function $F$, and we find in each case
\begin{equation}
\begin{aligned}\label{eq:Frhosol}
F^{L}(\rho)=&\left(\sinh(2\rho)\right)^{-i L_{\star}p_{+}} \left(\tanh(\rho)\right)^{-i L_{\star}p_{-}}\, ,\\
F^{R}(\rho)=&\left(\sinh(2\rho)\right)^{-i L_{\star}p_{-}} \left(\tanh(\rho)\right)^{-i L_{\star}p_{+}}\, .
\end{aligned}
\end{equation}
Then, it turns out that when one inserts these expressions into the equation of motion \eqref{eq:massiveeq}, one simply finds an algebraic equation either for $p_{-}$ or $p_{+}$, 
\begin{equation}
\begin{aligned}\label{eq:p1p2eq}
(\bar{\dal}-m_s^2)h_{(0)}^{L}=0\, &\Rightarrow\, 4 L_{\star} p_{+}^2+4 i p_{+}+L_{\star}m_s^2=0\, ,\\
(\bar{\dal}-m_s^2)h_{(0)}^{R}=0\, &\Rightarrow\, 4 L_{\star} p_{-}^2+4 i p_{-}+L_{\star}m_s^2=0\, .
\end{aligned}
\end{equation}
Each of these equations has the following solutions
\begin{align}\label{eq:p1p2sol}
h_{(0)}^{L}\Rightarrow\, p_{+}=-\frac{i}{L_{\star}}\mathfrak{h}^{s}_{\pm}(m_s^2)\, ,\quad h_{(0)}^{R}\Rightarrow\,p_{-}=-\frac{i}{L_{\star}}\mathfrak{h}^{s}_{\pm}(m_s^2)\, ,
\end{align}
where
\begin{equation}\label{eq:hs}
2\mathfrak{h}^{s}_{\pm}(m_s^2)\equiv 1\pm \sqrt{1+L_{\star}^2m_s^2}\, ,
\end{equation}
are, from an holographic perspective, the conformal dimensions of each of the modes of the scalar field. Now, with this result the asymptotic behavior of $h_{(0)}^{L,R}$ is 
\begin{equation}
h_{(0)}^{L,R}\sim e^{-2\rho\, \mathfrak{h}^{s}_{\pm}(m_s^2)}\, ,
\end{equation}
and, since we are only interested in solutions in which only the normalizable mode is active, we choose the ones with $\mathfrak{h}_{\pm}(m_s^2)>0$. When $m_s^2>0$ this means that we only keep the solution with $\mathfrak{h}^{s}_{+}(m_s^2)$. However, for $-1<L_{\star}^2m_s^2<0$, the mode $\mathfrak{h}^{s}_{-}(m_s^2)$ is also normalizable. 

Now, let us introduce the frequency and the momentum of these modes,
\begin{equation}\label{eq:omegapp}
\omega=p_{+}+p_{-}\, ,\quad m=L_{\star}(p_{-}-p_{+})\, ,
\end{equation}
where $m\in \mathbb{Z}$ is an integer, since we take $\phi$ to have a periodicity of $2\pi$. We also introduce the tortoise coordinate 
\begin{equation}
\rho_{*}=\int\frac{\diff \rho L_{\star}}{\sinh(\rho)}=L_{\star}\log\left[\tanh(\rho/2)\right]\, ,
\end{equation}
and in terms of this, we see that near the horizon $\rho_{*}\rightarrow -\infty$, these solutions behave as
\begin{equation}
h_{(0)}^{L,R}\sim e^{-i\omega(t+\rho_{*})+i m\phi}\, .
\end{equation}
This corresponds to waves moving towards the horizon, and hence we have shown that these solutions are quasinormal modes.  Their associated quasinormal mode frequencies (QNFs) are obtained from eqs.~\eqref{eq:omegapp} and \eqref{eq:p1p2sol}, and they read
\begin{equation}
\omega_{s}^{(0)}L_{\star}=\pm m-2i\, \mathfrak{h}^{s}_{+}(m_s^2)\, ,
\end{equation}
where $+m$ is for the left-moving modes and $-m$ for the right-moving ones. 

Finally, as shown in~\cite{Sachs:2008gt},  one can generate the infinite set of overtones from these fundamental quasinormal modes by applying the combination $L_{-1}\bar{L}_{-1}$, namely
\begin{equation}
h_{(n)}^{L,R}=(L_{-1}\bar{L}_{-1})^{n}h_{(0)}^{L,R}\, .
\end{equation}
These are also QNMs, because they satisfy the appropriate boundary conditions, and the effect of the operation $L_{-1}\bar{L}_{-1}$ is to shift $\omega\rightarrow \omega-2i/L_{\star}$. Therefore, the complete set of QNFs for the scalar field is
\begin{equation}
\omega_{s}^{(n)}L_{\star}=\pm m-2i\,\left(n+ \mathfrak{h}^{s}_{+}(m_s^2)\right)\, .
\end{equation}

Let us now move to the case of the massive graviton. The quasinormal modes of this field were computed in \cite{Myung:2011bn} in the context of New Massive Gravity, by using the fact that the equation \eqref{eq:massiveeq} can be written as the square of the equation for a chiral massive graviton, and hence one can apply the results of \cite{Sachs:2008gt} for Topologically Massive Gravity \cite{Deser:1981wh,Deser:1982vy}. Here we follow a direct analysis of \req{eq:massiveeq}. 

We need to find a symmetric, transverse and traceless field $\hat{h}_{ab}$ that satisfies \req{eq:massiveeq} with the QNM boundary conditions. Again, we can apply the technique of \cite{Sachs:2008gt} and find the chiral highest weight modes, which by definition satisfy
\begin{equation}\label{eq:L1massive}
L_{1}\hat{h}_{(0)ab}^{L}=0\, ,\qquad \bar{L}_{1}\hat{h}_{(0)ab}^{R}=0\, .
\end{equation}
It turns out that an appropriate ansatz for these modes is the following,
\begin{align}
\hat{h}_{(0)ab}^{L}\diff x^{a}\diff x^{b}&=F^{L}(\rho)e^{-i u p_{+}-i v p_{-}}\left(\diff v+\frac{2L_{\star}\diff \rho}{\sinh(2\rho)}\right)^2\, ,\\
\hat{h}_{(0)ab}^{R}\diff x^{a}\diff x^{b}&=F^{R}(\rho)e^{-i u p_{+}-i v p_{-}}\left(\diff u+\frac{2L_{\star}\diff \rho}{\sinh(2\rho)}\right)^2\, .
\end{align}
These are traceless and, interestingly, the conditions in \req{eq:L1massive} are equivalent to imposing them to be transverse, $\bar\nabla^{a}\hat{h}_{ab}=0$. Thus, we find the following first-order equations for the functions $F^{L,R}(\rho)$, 
\begin{equation}
\begin{aligned}
\left(\frac{2iL_{\star}(p_{-}+p_{+}\cosh(2\rho))}{\sinh(2\rho)}+\partial_{\rho}\right)F^{L}&=0\, ,\\
\left(\frac{2iL_{\star}(p_{+}+p_{-}\cosh(2\rho))}{\sinh(2\rho)}+\partial_{\rho}\right)F^{R}&=0\, .
\end{aligned}
\end{equation}
These are the same equations as those for the radial functions in the scalar case, and therefore the solutions are given by \req{eq:Frhosol}.
Then, we have to solve the equations of motion \eqref{eq:massiveeq}, and, again, when we insert the previous result, we see that all the components of the equations are reduced to an algebraic equation either for $p_{+}$ or $p_{-}$, 
\begin{equation}
4 L_{\star} p_{\pm}^2-4 i p_{\pm}+L_{\star}m_g^2=0\, .
\end{equation}
The solutions in this case are:
\begin{align}\label{eq:p1p2solg}
\hat{h}_{(0)ab}^{L}\Rightarrow\, p_{+}=-\frac{i}{L_{\star}}\mathfrak{h}^{g}_{\pm}(m_g^2)\, ,\quad \hat{h}_{(0)ab}^{R}\Rightarrow\,p_{-}=-\frac{i}{L_{\star}}\mathfrak{h}^{g}_{\pm}(m_g^2)\, ,
\end{align}
where
\begin{equation}
\mathfrak{h}^{g}_{\pm}(m_g^2)=\frac{-1\pm \sqrt{1+L_{\star}^2m_g^2}}{2}\, .
\end{equation}
Note the $-1$ rather than the $+1$ with respect to $\mathfrak{h}^{s}_{\pm}(m_g^2)$ in \req{eq:hs}. 
As in the case of the scalar field, it is immediate to verify that all these modes behave as outgoing waves at the horizon (they fall towards the horizon), but at infinity only the modes associated to $\mathfrak{h}^{g}_{+}(m_g^2)$ are normalizable, and hence only these are QNMs. By using again \eqref{eq:omegapp}, we obtain the frequencies of these fundamental modes
\begin{equation}
\omega_{g}^{(0)}L_{\star}=\pm m-2i\mathfrak{h}^{g}_{+}(m_g^2)\, ,
\end{equation}
where the $+m$ case is for left movers and $-m$ for the right movers. 
We note that, in this case, if $m_g^2<0$, the imaginary part becomes positive and the modes become unstable, unlike in the case of scalar perturbations. 
Finally, the overtones can be obtained by applying the operator $L_{-1}\bar{L}_{-1}$, which shifts the imaginary part of the frequency in $-2/{L_{\star}}$, as happened for the scalar. Therefore, the complete set of quasinormal mode frequencies for the massive graviton is
\begin{equation}
\omega_{g}^{(n)}L_{\star}=\pm m-2i\,\left(n+ \mathfrak{h}^{g}_{+}(m_g^2)\right)\, .
\end{equation}

\subsubsection{General BTZ}
The results from the previous subsubsection can be easily generalized to the BTZ black hole with arbitrary mass and angular momentum by noticing that the metric \eqref{eq:rotatingBTZ} can be mapped to eq. \eqref{eq:unitBTZ}. Starting with the rotating BTZ black hole \eqref{eq:rotatingBTZ}, we can perform the change of variables
\begin{equation}
t=\frac{r_{+}L_{\star}}{r_{+}^2-r_{-}^2}\left(\tilde{t}+\frac{r_{-}L_{\star}}{r_{+}}\tilde{\phi}\right)\, ,\quad
\phi=\frac{r_{+}L_{\star}}{r_{+}^2-r_{-}^2}\left(\tilde{\phi}+\frac{r_{-}}{r_{+}L_{\star}}\tilde{t}\right)\, \quad
r=\sqrt{r_{-}^2+(r_{+}^2-r_{-}^2)\cosh^2(\rho)}\, ,
\end{equation}
which leads to the metric 
\begin{equation}\label{eq:unitBTZ3}
\diff s^2=-\sinh^2(\rho) \diff \tilde{t}^2+L_{\star}^2\cosh^2(\rho) \diff \tilde{\phi}^2+L_{\star}^2\diff \rho^2\, ,
\end{equation}
which is locally the same as eq. \eqref{eq:unitBTZ2}. However, the geometry is different, because the ranges of the coordinates are different. In particular, $\tilde\phi$ is not to be considered an angular coordinate. 
Expressed in this way, we can study the perturbations as we did in the previous subsection by introducing the coordinates $\tilde{u}=\tilde{t}+L_{\star}\tilde\phi$, $\tilde v=\tilde t-L_{\star}\tilde\phi$. Then, we search for solutions with a dependence $\sim e^{-i\tilde u p_{+}-i\tilde v p_{-}}$, and the result of this analysis is the one we have just seen: either $p_{+}$ (for left-moving modes) or $p_{-}$ (for right-moving ones) must have a certain value for quasinormal modes. This is given by \req{eq:p1p2sol} for the scalar mode and by \req{eq:p1p2solg} for the massive graviton. 
However, now the frequency and the momentum are  identified by $-i\tilde u p_{+}-i\tilde v p_{-}=-i\omega t+i m\phi$, where $m$ is an integer. This yields the relations
\begin{align}
\omega L_{\star}&=p_{+}(r_{+}-r_{-})+p_{-}(r_{+}+r_{-})\, ,\\
m&=p_{-}(r_{+}+r_{-})-p_{+}(r_{+}-r_{-})\, .
\end{align}
Inserting the appropriate value of $p_{+}$ or $p_{-}$ in these equations, we get the QNFs as a function of the momentum. In full detail, we obtain the following set of frequencies
\begin{align}
\omega_{s}^{L,R}=&\pm \frac{m}{L_{\star}}-i\frac{r_{+}\mp r_{-}}{L_{\star}^2}\left(2n+1+ \sqrt{1+L_{\star}^2m_s^2}\right)\, ,\\
\omega_{g}^{L,R}=&\pm \frac{m}{L_{\star}}-i\frac{r_{+}\mp r_{-}}{L_{\star}^2}\left(2n-1+ \sqrt{1+L_{\star}^2m_g^2}\right)\, ,
\end{align}
with $n=0,1,\dots$ Additionally, recall that if $-1\le L_{\star}^2m_{s}^2\le 0$, there is a second family of scalar modes, with $-\sqrt{1+L_{\star}^2m_s^2}$.
From the point of view of holography, these results would be interpreted as the poles of the retarded Green functions of a thermal two-dimensional CFT placed on a circle of radius $L_{\star}$.  This can be generalized to a circle of arbitrary radius $R$ by performing a rescaling of the boundary metric, identifying $\omega'=\omega L_{\star} /R$ and $T'=T L_{\star}/R$. Now, the left- and right-moving Virasoro algebras of the $1+1$ dimensional CFT split the theory in two sectors, that in the background of the BTZ black hole have temperatures \cite{Birmingham:2001pj}
\begin{equation}
T^{L}=\frac{r_{+}-r_{-}}{2\pi L_{\star}^2}\, ,\quad T^{R}=\frac{r_{+}+r_{-}}{2\pi L_{\star}^2}\, .
\end{equation}
We can then write the QNFs as 
\begin{align}
\omega_{s}^{L,R}=&\pm \frac{m}{R}-2i\pi T^{L,R}\left(2n+1+ \sqrt{1+L_{\star}^2m_s^2}\right)\, ,\\
\omega_{g}^{L,R}=&\pm \frac{m}{R}-2i\pi T^{L,R}\left(2n-1+ \sqrt{1+L_{\star}^2m_g^2}\right)\, .
\label{eq:massiveomega}
\end{align}
As noted in \cite{Birmingham:2001pj}, the scalar QNFs precisely match the field theory computation of the poles of retarded thermal correlators, which is considered to be a remarkable test of the AdS/CFT correspondence. Holography tells us that \req{eq:massiveomega} should compute the poles of the thermal correlators for a CFT dual to a bulk theory with a massive graviton.

\section{Theories satisfying an holographic c-theorem}\label{ctheorem}
Interesting extensions of Einstein and New Massive Gravities to higher orders can be obtained by demanding that the corresponding densities satisfy a simple holographic c-theorem \cite{Sinha:2010ai,Paulos:2010ke}. This set of theories is defined by the property that they yield second-order equations when evaluated on the ansatz
\begin{equation}\label{eq:RGmetric}
\diff s^2=\diff \rho^2+a(\rho)^2[-\diff t^2+\diff x^2]\,.
\end{equation}
Supplementing the action with an appropriate stress-tensor, the metric can be made to interpolate between  two asymptotic AdS$_3$ regions \cite{Freedman:1999gp,Girardello:1998pd} which, from the CFT point of view would represent IR and UV fixed points. Intermediate values of the holographic coordinate are then interpreted as representing the RG flow between both CFTs.


The idea behind the holographic c-theorem\footnote{The c-theorem for general $2d$ CFTs has been proven in \cite{Zamolodchikov:1986gt,Casini:2017vbe}.} involves constructing a function $c(\rho)$ which decreases monotonously along the RG flow, as we move from the UV to the IR. In the present holographic context, the fixed points can be chosen to be  $\rho_{\rm UV}= +\infty$  and   $\rho_{\rm IR}=-\infty$, 
so a function satisfying
\begin{equation}
    c'(\rho) \geq 0  \quad \forall\, \rho\, ,
\end{equation}
does the job. 
Now,  the usual holographic c-theorem construction involves considering a function $c(\rho)$ such  that  $c'(\rho)$ is  proportional to the combination of stress-tensor components $T_t^t-T_{\rho}^{\rho}$. Then, imposing that the stress-tensor satisfies  the null energy   condition, such combination has a sign, namely,
\begin{equation}
    T_t^t-T_{\rho}^{\rho}  \overset{\rm  \ssc NEC}{\leq} 0\, .
\end{equation}
Therefore, any $c(\rho)$ such that  $c'(\rho)\propto -(T_t^t-T_{\rho}^{\rho})$, up to an overall positive-definite constant, satisfies the requirement.

For theories of the type considered above, it is straightforward to construct an appropriate c-function such  that \cite{Freedman:1999gp,Myers:2010xs,Myers:2010tj} 
\begin{equation}
    c'(\rho)=-\frac{a^2}{8G a'^{2}}\,[T_t^t-T_{\rho}^{\rho}]\, .
\end{equation}
This can be obtained from the Wald-like \cite{Wald:1993nt} formula \cite{Sinha:2010ai,Myers:2010tj}
\begin{equation}
    c(\rho)\equiv \frac{\pi a}{2 a'}\frac{\partial \mathcal{L}}{\partial R^{t\rho}\,_{t\rho}}\, ,
\end{equation}
where the Lagrangian derivative  components are evaluated on \req{eq:RGmetric}. By construction, $c(\rho)$ coincides with the Virasoro central charges of the fixed-point theories.

As argued in \cite{Paulos:2010ke}, demanding second-order equations for the ansatz \req{eq:RGmetric} for a set of order-$n$ densities amounts at imposing $n-1$ conditions. The idea is to consider the on-shell evaluation of the corresponding Lagrangian densities and impose that neither terms involving derivatives of $a(\rho)$ higher than two, nor powers of  $a''(\rho)$ higher than one appear in the resulting expression. This enforces the corresponding equations of motion to be second-order and that a simple c-function can be defined from the above formulas.

As we have shown, there are $\#(n)$ independent densities at order $n$, which means that there are $\#(n)-(n-1)$ independent order-$n$ densities which satisfy a simple holographic c-theorem. Hence, for $n=1,\dots,5$, there is a single such density at each order, but degeneracies start to appear at order six. As observed in \cite{Paulos:2010ke}, it is always possible to write the corresponding linear combination of order-$n$ densities satisfying a simple holographic c-theorem as a single density which has a non-trivial on-shell action when evaluated on \req{eq:RGmetric}, plus densities which simply vanish when evaluated on such ansatz. Hence, we learn that there are $\#(n)-n$ independent  order-$n$ densities which are trivial on the \req{eq:RGmetric} ansatz. Remarkably, as we show below, all such densities of arbitrary orders turn out to be proportional to a single sextic density which identically vanishes on the metric \eqref{eq:RGmetric}. As for the densities which contribute non-trivially to the holographic c-function we find a new recursive formula which allows for the construction of the corresponding order-$n$ density from the order-$(n-1)$, the order-$(n-2)$, the Einstein gravity and the NMG densities. The recurrence can be solved explicitly, and so we are able to provide an explicit formula for a general order density which non-trivially satisfies the holographic c-theorem. Finally, we explore the relation between such general order density and the one resulting from the expansion of previously proposed Born-Infeld gravities which also satisfy the holographic c-theorem. Naturally, the relation always involves densities trivially satisfying the holographic c-theorem.  



\subsection{Recursive formula}



As we have mentioned, at each order there is a single possible functional dependence on $a(\rho)$ of the on-shell action of theories satisfying the holographic c-theorem. Then, up to terms which do not contribute when evaluated on \req{eq:RGmetric}, there is a unique  such density at each curvature order. The on-shell expressions for $R,\mathcal{S}_2,\mathcal{S}_3$ read
\begin{align}
   \left. R\right|_a  
    = -\frac{2(a'^2 +2 a a'') }{ a^2} \, , \quad
  \left.  \mathcal{S}_2\right|_a  
    = \frac{2(a'^2 - a a'')^2}{3 a^4} \, , \quad
  \left.  \mathcal{S}_3\right|_a  
    = \frac{2(a'^2 - a a'')^3}{9 a^6}\, .
\end{align}
As observed in \cite{Paulos:2010ke}, the on-shell Lagrangian of densities satisfying the holographic c-theorem in a non-trivial fashion follows the simple pattern 
\begin{equation}\label{eq:Lnonshell}
\left. \mathcal{C}_{(n)}\right|_{a}=\left(\frac{a'}{a}\right)^{2(n-1)}\left[\frac{a''}{a}+\frac{3-2n}{2n}\left(\frac{a'}{a}\right)^{2}\right]\, .
\end{equation}
With this choice of normalization, the first three densities read
\begin{align}
\mathcal{C}_{(1)}&=-\frac{1}{4}R\, ,\\
\mathcal{C}_{(2)}&=+\frac{3 R^2}{16}-\frac{\mathcal{R}_2}{2}\\ &=+\frac{R^2}{48}-\frac{\mathcal{S}_2}{2}\, ,\\
\mathcal{C}_{(3)}&=-\frac{17 R^3}{48}+\frac{3 R \mathcal{R}_2}{2}-\frac{4 \mathcal{R}_3}{3}\\ &=-\frac{R^3}{432}+\frac{R \mathcal{S}_2}{6}-\frac{4 \mathcal{S}_3}{3}\, .
\end{align}
Now, an easy way to prove that instances of non-trivial densities actually exist at arbitrarily high orders is by finding a recursive relation.
Since, essentially, these densities are defined by the form of their on-shell Lagrangian on the RG-flow metric (\ref{eq:RGmetric}), we can try to derive such recursive relations by using \req{eq:Lnonshell}. We find the particularly simple relation,
\begin{equation}\label{recuu}
\mathcal{C}_{(n)}=\frac{4(n-1)(n-2)}{3n(n-3)}\left(\mathcal{C}_{(n-1)}\mathcal{C}_{(1)}-\mathcal{C}_{(n-2)}\mathcal{C}_{(2)}\right)\, .
\end{equation} 
This expression allows us to generate holographic $c$-theorem satisfying densities of arbitrary orders once we know $\mathcal{C}_{(1)}$, $\mathcal{C}_{(2)}$ and $\mathcal{C}_{(3)}$, which are given above. Since $\mathcal{C}_{(4)}$ and $\mathcal{C}_{(5)}$ are unique, this formula should give precisely those densities. This is indeed the case, and one finds
\begin{align}
    \mathcal{C}_{(4)} & =+ \frac{41R^4}{384} - \frac{3R^2\mathcal{R}_2}{8} + \frac{2R\mathcal{R}_3}{3} - \frac{\mathcal{R}_2^2}{2}
    \\ &=+ \frac{R^4}{3456} - \frac{R^2\mathcal{S}_2}{24} + \frac{2R\mathcal{S}_3}{3} - \frac{\mathcal{S}_2^2}{2},
\end{align}
and
\begin{align}
    \mathcal{C}_{(5)} & =+ \frac{61R^5}{960} - \frac{7R^3\mathcal{R}_2}{12} + \frac{2R^2\mathcal{R}_3}{15} + \frac{7R\mathcal{R}_2^2}{5} - \frac{16\mathcal{R}_2\mathcal{R}_3}{15}
    \\ & = -\frac{R^5}{25920} + \frac{R^3\mathcal{S}_2}{108} - \frac{2R^2\mathcal{S}_3}{9} + \frac{R\mathcal{S}_2^2}{3} - \frac{16\mathcal{S}_2\mathcal{S}_3}{15},
\end{align}
which agree with the results previously reported in \cite{Sinha:2010ai,Paulos:2010ke}. On the other hand, for $n\ge 6$ the recursion produces a single representative non-trivial density. For example, for $n=6$ ---which is the order at which degeneracies start to appear due to the existence of densities trivially satisfying the holographic c-theorem--- we find from the recursive formula
\begin{align}
    \mathcal{C}_{(6)} & = -\frac{1103 R^6}{20736} + \frac{115R^4\mathcal{R}_2}{288} - \frac{19R^3\mathcal{R}_3}{81} - \frac{71R^2\mathcal{R}_2^2}{108} + \frac{8R\mathcal{R}_2\mathcal{R}_3}{9} - \frac{10 \mathcal{R}_2^3}{27}
   \\ & = +\frac{R^6}{186624} - \frac{5R^4\mathcal{S}_2}{2592} + \frac{5R^3\mathcal{S}_3}{81} - \frac{5R^2\mathcal{S}_2^2}{36} + \frac{8R\mathcal{S}_2\mathcal{S}_3}{9} - \frac{10 \mathcal{S}_2^3}{27}\, .
\end{align}

\subsection{General formula for order-$n$ densities}
Interestingly, it is possible to solve the two-term recurrence relation (\ref{recuu}) analytically and obtain an explicit expression for the order-$n$ density non-trivially satisfying the holographic c-theorem. The result  which, once again, takes a simpler form in terms of the $\{R,\mathcal{S}_{2},\mathcal{S}_3\}$ set, reads,
\begin{equation}\label{eq:cdensityn}
\begin{aligned}
\mathcal{C}_{(n)}=\frac{3 (-1)^{n}}{4\cdot 6^n n}\Bigg\{&\left(R+\sqrt{24\mathcal{S}_2}\right)^{n-1}\left(R-(n-1)\sqrt{24\mathcal{S}_2}\right)\left(1-\sqrt{6}\frac{\mathcal{S}_3}{\mathcal{S}_2^{3/2}}\right)\\
+&\left(R-\sqrt{24\mathcal{S}_2}\right)^{n-1}\left(R+(n-1)\sqrt{24\mathcal{S}_2}\right)\left(1+\sqrt{6}\frac{\mathcal{S}_3}{\mathcal{S}_2^{3/2}}\right)
\Bigg\}\, .
\end{aligned}
\end{equation}
Even though this expression may look odd because it depends in a non-polynomial way on the densities, it does reduce to a polynomial expression when we evaluate it for any integer $n\ge 1$. One can check this by expanding the $\left(R\pm \sqrt{24\mathcal{S}_2}\right)^{n-1}$ terms using the binomial coefficients.  In particular, note that this formula is even under the exchange $\mathcal{S}_2^{1/2}\rightarrow- \mathcal{S}_2^{1/2}$, and therefore $\mathcal{S}_2^{1/2}$ always appears with even powers (\textit{i.e.}, there are no square roots). Explicitly, the result of this expansion reads
\begin{equation}\label{CnFormula}
\begin{aligned}
\mathcal{C}_{(n)}=\frac{3 (-1)^{n}}{2\cdot 6^n n}\Bigg\{&\sum_{k=0}^{\lfloor \frac{n}{2} \rfloor}(24 \mathcal{S}_2)^k R^{n-2k}\left[\binom{n-1}{2k}-(n-1)\binom{n-1}{2k-1}\right]\\
&-288\mathcal{S}_3\sum_{k=0}^{\lfloor\frac{n-3}{2}\rfloor}(24 \mathcal{S}_2)^k R^{n-3-2k}\left[\binom{n-1}{2k+3}-(n-1)\binom{n-1}{2k+2}\right]\Bigg\}\, ,
\end{aligned}
\end{equation}
which is valid whenever $n\in \mathbb{N}$. 

Interestingly, the density \eqref{eq:cdensityn} can also be applied for non-integer $n$, since it always yields the result \eqref{eq:Lnonshell} when evaluated on the metric \eqref{eq:RGmetric}, and therefore it yields second-order equations for the RG-flow metric.  Hence, these Lagrangians provide a generalization of the  holographic $c$-theorem-satisfying densities for arbitrary real values of $n$. 

\subsection{All densities with a trivial c-function emanate from a single sextic density }
For the first five curvature orders, there exists a single density which satisfies the holographic c-theorem condition. Now, for $n=6$, there exists an additional density,
\begin{align}\label{Omeg6}
\Omega_{(6)} &\equiv 6 \mathcal{S}_3 ^2-\mathcal{S}_2 ^3\\ &=\frac{1}{3} \left[R^6-9R^4\mathcal{R}_2+8R^3\mathcal{R}_3+21R^2\mathcal{R}_2^2-36R\mathcal{R}_2\mathcal{R}_3-3\mathcal{R}_2^3+18\mathcal{R}_3^2\right]\, ,
\end{align}
with the property of being identically vanishing when evaluated on the c-theorem ansatz (\ref{eq:RGmetric}) and which therefore does not
contribute to the equations of motion for that ansatz. 

An immediate consequence is that any product of $\Omega_{(6)} $ with any other density also satisfies trivially the holographic c-theorem.
 Therefore, for $n\ge 6$ we have, at least, the following set of densities which satisfy the holographic c-theorem
\begin{equation}\label{cThLs}
\mathcal{L}_{(n)}^{\rm c-theorem}=\alpha_n \mathcal{C}_{(n)}+\Omega_{(6)}\cdot \mathcal{L}^{\rm general}_{(n-6)}\, ,
\end{equation}
where $\mathcal{L}^{\rm general}_{(n-6)}$ is the general Lagrangian of order $n-6$ in the curvature. Remarkably, these are all the densities of this type that exist. 

This can be proven as follows. First, observe that there exist $\#(n-6)$ densities of order $n-6$. Hence, there exists the same number of order-$n$ densities in the set $\Omega_{(6)} \cdot \mathcal{L}^{\rm general}_{(n-6)}$. Now, as observed earlier, there exist $\#(n)-(n-1) $ independent order-$n$ densities which satisfy the holographic c-theorem, one of which does so in a non-trivial fashion. The latter can be chosen to be $\mathcal{C}_{(n)}$ and we are left with $\#(n)-n$ independent densities which trivially satisfy the holographic c-theorem. Now, invoking the result in \req{properttt}, we observe that this number exactly matches the number of densities in the set $\Omega_{(6)} \cdot \mathcal{L}^{\rm general}_{(n-6)}$.

 In sum, $\mathcal{L}_{(n)}^{\rm c-theorem}$ as defined above is the most general higher-curvature order-$n$ density satisfying the holographic c-theorem and all densities satisfying it in a trivial fashion emanate from the sextic density $\Omega_{(6)}$. This is a rather intriguing result which suggests that there may be something more fundamentally special about this density. As a matter of fact, this will not be the last time we encounter it.


\subsection{Absence of scalar mode in the spectrum}
An immediate consequence of \req{cThLs} is that none of the densities trivially satisfying the holographic c-theorem contributes to the linearized equations around an Einstein metric. This is because all densities involved take the form $\Omega_{(6)} \cdot \mathcal{L}^{\rm general}_{(n-6)}$  and therefore belong to the set $\mathcal{G}|_{\text{trivial linearized equations}}$ as defined in \req{gtriv}. On the other hand, we can use our previous results to prove that densities which satisfy the holographic c-theorem in a non-trivial fashion do not incorporate the scalar mode in their spectrum. This latter property seems to have been observed in certain particular cases \cite{Afshar:2014ffa} but we have found no general proof in the literature.

We saw in section \ref{seclineq} that the condition for the absence of the scalar mode in the linearized spectrum, $m_s^2 \to \infty$, was satisfied by theories of the form
\begin{equation}
\mathcal{G}(R,\mathcal{S}_2,\mathcal{S}_3) = f(R) + \mathcal{S}_2 g(R) + \mathcal{G}_{\rm triv}
\end{equation}
for which
\begin{equation}\label{12frr}
    12 \bar{f}_{R,R} + \bar{g} = 0.
\end{equation}

For theories where $\mathcal{G}(R,\mathcal{S}_2,\mathcal{S}_3)$ is a polynomial, as the ones we are considering, this cancellation must occur order by order. At any given order $n$ the only possible forms of $f$ and $g$ are $f_{(n)}(R) = \lambda_{(n)} R^n$ and $g_{(n)}(R) = \mu_{(n)}R^{n-2}$ for some constants $ \lambda_{(n)}$ and $ \mu_{(n)}$, and so $12 \bar{f}_{R,R} + \bar{g} = \left[12n(n-1)\lambda_{(n)} + \mu_{(n)} \right] \bar{R}^{n-2}$, and so condition (\ref{12frr}) becomes
\begin{equation}\label{cTnoS}
    12n(n-1)\lambda_{(n)} + \mu_{(n)} = 0,
\end{equation}
where $\lambda_{(n)} = \beta_{n00}$ is the coefficient in front of the $R^n$ term and $\mu_{(n)} = \beta_{(n-2)10}$ is the coefficient in front of the $\mathcal{S}_2R^{n-2}$ term.


Now, expanding eq. \eqref{CnFormula} and keeping only the terms with $k = 0,1$ in the first sum, we see
\begin{equation}
    \mathcal{C}_{(n)} = \frac{3(-1)^n}{2\cdot6^nn} \Big\{ R^n - 12n(n-1) \mathcal{S}_2 R^{n-2} + \cdots \Big\},
\end{equation}
and so 
\begin{equation}
    \lambda_{(n)} = \frac{3(-1)^n}{2\cdot6^nn}, \quad \quad \mu_{(n)} = - 12n(n-1)\frac{3(-1)^n}{2\cdot6^nn},
\end{equation}
which clearly fulfill condition \eqref{cTnoS}. This proves that all theories satisfying the holographic c-theorem have a linearized spectrum which does not include the scalar mode. 



\subsection{Born-Infeld gravity }

It was proposed in \cite{Gullu:2010pc} that New Massive Gravity could also be extended through a Born-Infeld gravity theory with Lagrangian density
\begin{equation}\label{BINMG}    \mathcal{L}_{\text{BI-NMG}} = \sqrt{\det \left( \delta_a^b + \frac{\sigma}{m^2} G_a^b \right) }  - \left( 1 - \frac{\Lambda}{2m^2} \right),
\end{equation}
where $G_{ab} = R_{ab} - \frac{1}{2}g_{ab}R$ is the Einstein tensor and $\sigma = \pm 1$. This theory reproduces NMG when expanded to quadratic order in the curvature. Then, after \cite{Sinha:2010ai} proved that both NMG and the cubic order term of eq. \eqref{BINMG} admitted an holographic c-function, it was soon proven in \cite{Gullu:2010st} that the full theory also satisfied a simple holographic c-theorem of the same kind as the one described in the previous subsections. The cancellations on the on-shell evaluation of these theories required by the c-theorem construction occur order by order, and so the theory defined by eq. \eqref{BINMG} generates an infinite number of higher derivative densities which non-trivially fulfil an holographic c-theorem at any truncated order \cite{Alkac:2018whk}.

Now, in view of our results, it would be interesting to know whether the terms generated by the expansion of \eqref{BINMG} order by order, which we shall call $\mathcal{B}_{(n)}$, are the same ones as the non-trivial densities $\mathcal{C}_{(n)}$ generated by the recursive formula \eqref{recuu}.
Following what we have just learned in the previous section, that should indeed be the case for $n = 1,\ldots,5$. For $n\geq6$ we expect both sets of densities to coincide up to ``trivial'' densities, and we find that to be the case.

Let us expand the density \eqref{BINMG}. We set $\sigma = 1$ and $m^2 = 1$ for simplicity, as they can be easily restored by dimensional analysis. In 3 dimensions the determinant of any matrix can be computed as 
\begin{equation}
    \det (A) = \frac{1}{6} \left[ \left( \tr (A) \right)^3 - 3 \tr (A) \tr (A^2) + 2 \tr(A^3) \right].
\end{equation}
In our case, we have $A = \mathbb{1} + g^{-1}G$, which gives
\begin{equation}
    \det (\mathbb{1} + g^{-1}G ) = 1 + \frac{-1}{2}R + \frac{1}{4} \mathcal{T}_2 + \frac{1}{24}\mathcal{T}_3
\end{equation}
where we have defined
\begin{align}
    \mathcal{T}_2 & \equiv R^2 - 2 \mathcal{R}_2 = \frac{1}{3} R^2 - 2 \mathcal{S}_2,
    \\
    \mathcal{T}_3 & \equiv  R^3 - 6 R \mathcal{R}_2 + 8 \mathcal{R}_3 = \frac{-1}{9}R^3 + 2 R \mathcal{S}_2 + 8 \mathcal{S}_3.
\end{align}
We can now simply Taylor expand the square root, $\sqrt{1+x} = \sum_{m=0}^{\infty} \binom{1/2}{m} x^m$, with $x = \det (\mathbb{1} + g^{-1}G )-1$ and then collect the relevant terms at each order $n$ to build $\mathcal{B}_{(n)}$. The result is the following,
\begin{equation}\label{Bn}
    \mathcal{B}_{(n)} = \hat{\sum} \binom{1/2}{i+j+k} \frac{(i+j+k)!}{i!j!k!}\left( \frac{-1}{2}R \right)^i\left( \frac{1}{4}\mathcal{T}_2 \right)^j \left( \frac{1}{24} \mathcal{T}_3 \right)^k ,
\end{equation}
where the sum is performed over the indices $i,j,k$ that fulfill the integer partition $n = i+2j+3k$.

The lowest order densities given by the formula above are
\begin{equation}
    \mathcal{B}_{(1)}  = \mathcal{C}_{(1)}\, ,\quad  
    \mathcal{B}_{(2)}   = \frac{1}{2}\mathcal{C}_{(2)} \, , \quad 
    \mathcal{B}_{(3)}  =  -\frac{1}{8}\mathcal{C}_{(3)}\, , \quad 
    \mathcal{B}_{(4)}  = \frac{1}{16}\mathcal{C}_{(4)}\, , \quad 
    \mathcal{B}_{(5)}  = -\frac{5}{128}\mathcal{C}_{(5)}\, , 
\end{equation}
which are indeed proportional to the densities $\mathcal{C}_{(n)}$ found previously through the recursion relation \eqref{recuu}, as expected.
At the next orders, however, eq. \eqref{Bn} gives a different non-trivial density than the one given by the recursion relation \eqref{recuu}. Following eq. \eqref{cThLs}, we see that the relationship between the densities $\mathcal{B}_{(n)}$ and $\mathcal{C}_{(n)}$ at orders $n \geq 6$ is given by
\begin{equation}
    \mathcal{B}_{(n)} = (-1)^n \frac{(2n-5)!!}{(2(n-1))!!} \mathcal{C}_{(n)} + \Omega_{(6)} \cdot \mathcal{L}_{(n-6)},
\end{equation}
for some particular densities $\mathcal{L}_{(n-6)}$.
For example,
\begin{align}
    \mathcal{B}_{(6)} & = \frac{7}{256} \mathcal{C}_{(6)} - \frac{1}{432} \Omega_{(6)}, \\
     \mathcal{B}_{(7)} & = -\frac{21}{1024} \mathcal{C}_{(7)} - \frac{R}{576} \Omega_{(6)}, \\
     \mathcal{B}_{(8)} & = \frac{33}{2048} \mathcal{C}_{(8)} - \frac{11R^2 + 24\mathcal{S}_2}{13824} \Omega_{(6)}.
\end{align}
Hence, both $\mathcal{C}_{(n)} $ and $\mathcal{B}_{(n)} $ provide sets of order-$n$ densities which non-trivially satisfy the holographic c-theorem. While the $\mathcal{C}_{(n)} $ are distinguished by the property of satisfying the simple recurrence relation (\ref{recuu}), the $\mathcal{B}_{(n)} $ have the property of corresponding to the general term in the expansion of the Born-Infeld theory (\ref{BINMG}). Both sets are equal up to terms which identically vanish in the holographic c-theorem ansatz which, as we have seen, are all proportional to the density $\Omega_{(6)}$.

Another Born-Infeld theory has been proposed as a non-minimal extension of NMG \cite{Alkac:2018whk}, with Lagrangian density
\begin{equation}\label{nMBI}
    \mathcal{L}_{\text{nM-BI}} = \sqrt{\det \left( \delta_a^b - \frac{2}{m^2} P_a^b + \frac{1}{m^4} P_a^c P_c^b \right) }  - \left( 1 - \frac{\Lambda}{2m^2} \right),
\end{equation}
where $P_a^b = R_a^b - \frac{1}{4}\delta_a^b R$ is the Schouten tensor. The full theory also allows for an holographic c-function. However, when expanded order by order using a similar method as the one described above, we see that it does not produce an infinite number of higher derivative densities which non-trivially fulfil an holographic c-theorem. At order $n=2$ and $n=3$ we obtain terms proportional to $\mathcal{C}_{(2)}$ and $\mathcal{C}_{(3)}$, as expected, but the terms with $n \geq 4$ all trivialize due to the Schouten identities described in section 2. 
Therefore, the density \eqref{nMBI} is equivalent to the much simpler density 
\begin{equation}\label{nMBIv2}
    \mathcal{L} = R - 2\Lambda + \frac{2}{m^2} \left( \mathcal{R}_2 - \frac{3}{8}R^2 \right) + \frac{1}{m^4} \left( \frac{17}{48} R^3 - \frac{3}{2}R \mathcal{R}_2 + \frac{4}{3} \mathcal{R}_3 \right).
\end{equation}


\section{Generalized Quasitopological gravities}\label{GQTss}
A different classification criterion which has attracted a lot of attention in higher dimensions entails considering higher-curvature theories which admit generalizations of the $D$-dimensional Schwarzschild black hole characterized by a single function, \ie satisfying $g_{tt}g_{rr}=-1$. Theories of this kind have been coined ``Generalized quasitopological gravities'' (GQTs) \cite{Hennigar:2017ego,PabloPablo3,Bueno:2019ycr}, and include Quasitopological \cite{Quasi2,Quasi,Dehghani:2011vu,Cisterna:2017umf,Myers:2010jv} and Lovelock gravities \cite{Lovelock1,Lovelock2} as particular cases. 


Given a $D$-dimensional higher-curvature gravity with Lagrangian density $\mathcal{L}(R_{abcd},g^{ef})$. Let  $L_{f}$ be the effective Lagrangian 
obtained from the evaluation of $\sqrt{|g|}\mathcal{L}$ in the ansatz
\begin{equation}\label{fmetric}
\diff s^2=f(r)\diff t^2+\frac{\diff r^2}{f(r)}+\diff r^2 \diff \Omega_{(D-2)}^2 \, .
\end{equation}
Then, we say that $\mathcal{L}$ is a GQT gravity if
\begin{equation}\label{eq:GQTcond}
\frac{\partial L_f}{\partial f}-\frac{\diff}{\diff r}\frac{ \partial L_f}{\partial f'}+\frac{\diff^2}{\diff r^2}\frac{\partial L_f}{\partial f''}=0\, ,
\end{equation} 
namely, if the Euler-Lagrange equation of $f(r)$ identically vanishes \cite{PabloPablo3}. This is equivalent to requiring that $L_f$ is a total derivative, \ie 
\begin{equation} \label{totder}
L_f= \frac{\diff F_0}{\diff r} \, , \quad \text{ for some function} \quad F_0\equiv F_0[r,f(r),f'(r)]\, .
\end{equation}
Theories satisfying these requirements satisfy a number of interesting properties, such as admitting non-hairy generalizations of the Schwarzschild AdS$_D$ solution characterized by a single function or possessing a linearized  spectrum around maximally symmetric backgrounds identical to the Einstein gravity one. For a more detailed summary of the properties satisfied by GQTs see \eg \cite{Bueno:2019ltp}. In the same reference it has been proven that any gravitational effective action in $D\geq 4$ can be mapped via field redefinitions to a GQT.


Here we are interested in exploring the possible existence of GQTs in three dimensions. In order to do that, we need to determine the set of densities for which eq. (\ref{eq:GQTcond}) holds, if any. As a first step, we need to evaluate our fundamental building-block densities on such ansatz. Defining the quantities
\begin{equation}
A\equiv \frac{f''}{2},\quad B\equiv -\frac{f'}{2r},\quad \psi\equiv \frac{1-f}{r^2},
\end{equation}
which are the only functional dependences on $f(r)$ appearing in the Riemann tensor, we find
 \begin{align}
 R\big|_f= -2A+4B\, , \quad  \mathcal{S}_2\big|_f =\frac{2}{3}(A+B)^2 \, , \quad \mathcal{S}_3\big|_f =\frac{2}{9}(A+B)^3 \, , 
 \end{align}
 and 
\begin{equation}
\mathcal{R}_2\big|_f=2(A^2-2A B+3B^2)\, , \quad \mathcal{R}_3\big|_f =-2(A^3-3A^2 B+3A B^2+5B^3) \, .
\end{equation}
We immediately observe the absence of $\psi$ in these expressions, which means that three-dimensional on-shell Lagrangians do not depend on the function $f(r)$ explicitly, but only on its first and second derivatives.
Hence, in this case the GQT condition \eqref{eq:GQTcond} becomes simpler, namely,
\begin{equation}\label{integ}
\frac{ \partial L_f}{\partial f'}=\frac{\diff}{\diff r}\frac{\partial L_f}{\partial f''}+c\, ,
\end{equation}
where $c$ is an integration constant.

Evaluating on-shell a general order-$n$ density in the $\{R,\mathcal{S}_2,\mathcal{S}_3\}$ basis, we find
\begin{align}
L_{(n),f} &=r L^{2(n-1)} \sum_{j,k} \beta_{n-2j-3k,j,k} \frac{(-1)^{n-2j-3k}}{6^j 36^k} \left[ f''+\frac{2f'}{r}\right]^{n-2j-3k} \left[f''-\frac{f'}{r}\right]^{2j+3k}\, , \\
& =r L^{2(n-1)} \sum_{j,k,l,m} c_{jklm} {f''}^{(l+m)} \left[\frac{f'}{r}\right]^{n-m-l}
\end{align}
where $L_{(n),f}\equiv \sqrt{g}\mathcal{L}_{(n)}|_{f}$ and where we used the binomial expansion twice in the second line and defined the constants
\begin{equation}
c_{jklm}\equiv \frac{ \beta_{n-2j-3k,j,k}}{6^{j+2k}}  (-1)^{n-m} 2^{n-2j-3k-l}\binom{n-2j-3k}{l}\binom{2j+3k}{m} \, .
\end{equation}
The combination $l+m$ takes integer values from $0$ to $n$, and hence $L_{(n),f}$ can be written as a linear combination of terms with different powers of  $f''$ taking such values. Now, in order for  $L_{(n),f}$ to be a total derivative as required by \req{totder}, we need to impose that all terms involving powers of $f''$ higher than one vanish. This implies imposing $n-1$ conditions on the coefficients $\beta_{n-2j-3k,j,k}$. Once this is done, we are left with
\begin{align}
L_{(n),f} = g_1[r,f'(r)] + g_2[r,f'(r)] f''(r)\, , 
\end{align}
where
\begin{equation}
g_1\equiv r L^{2(n-1)} \sum_{j,k}c_{jk00} \left[\frac{f'}{r}\right]^{n}\, , \quad g_2\equiv r L^{2(n-1)} \sum_{j,k}(c_{jk10} +c_{jk01})  \left[\frac{f'}{r}\right]^{n-1}\, .
\end{equation}
However, the fact that $L_{(n),f}$ is linear in $f''(r)$ does not guarantee that $L_{(n),f}$ is a total derivative. In order for this to be the case, we need to impose the additional condition  given by \req{integ} which, in terms of $g_1$ and $g_2$ becomes 
\begin{equation}
\frac{\partial g_1}{\partial f'}=\frac{\partial g_2}{\partial r}\, .
\end{equation}
Explicitly, this condition becomes
\begin{equation}\label{condime}
n \sum_{j,k}c_{jk00} = (n-2) \sum_{j,k} (c_{jk10}+c_{jk01})\, ,
\end{equation}
which in terms of the original $\beta_{ijk}$ coefficients reads,
\begin{equation}
\sum_{j,k} \frac{\beta_{n-2j-3k,j,k}}{6^{j+2k}} 2^{n-2j-3k-1}[2-n+6j+9k]=0\, .
\end{equation}
Adding this to the $n-1$ conditions imposed earlier, we find a total of $n$ conditions to be imposed to $L_{(n),f} $ in order for it to be a GQT density. Hence, we have $\#(n)-n=\#(n-6)$ GQT densities at order $n$.   


\subsection{All GQT densities emanate from the same sextic density }
 Interestingly, the number of order-$n$ GQT densities exactly coincides with the number of densities trivially satisfying the holographic c-theorem. More remarkably, the two sets of densities are in fact identical. Indeed, the special sextic density $\Omega_{(6)}$ defined in \req{Omeg6} as the source of all densities trivially satisfying the holographic c-theorem turns out to be also the source of all GQT densities. Indeed, it is not difficult to see that 
 \begin{equation}
\left.  \Omega_{(6)}\right|_f=0\, ,
 \end{equation}
 which means that all densities involving $\Omega_{(6)}$ identically vanish and are therefore ``trivial'' GQT densities ---in the sense that they make no contribution to the equation of $f(r)$. Since there are $\#(n-6)$ of such densities, we learn that in fact all GQT densities in three dimensions are ``trivial'' and proportional to $\Omega_{(6)}$, 
  \begin{equation}
\mathcal{L}_{(n)}^{\rm GQT}=\Omega_{(6)}\cdot \mathcal{L}^{\rm general}_{(n-6)}\, ,
\end{equation}
where $\mathcal{L}^{\rm general}_{(n-6)}$ is the most general order-$(n-6)$ density.

In sum, we learn that in three dimensions there  exist no non-trivial GQTs. This situation is very different from higher-dimensions: in $D=4$ there exists one independent non-trivial GQT density for every $n\geq 3$ whereas for $D\geq 5$ there actually exist $n-1$ independent inequivalent  GQT densities for every $n$ ---namely, there exist $n-1$ densities of order $n$ each of which makes a functionally different contribution to the equation of $f(r)$ \cite{Buenooooo}. As a matter of fact, the triviality of the three-dimensional case unveiled here is not so surprising given that all higher-curvature theories admit the BTZ solution ---as opposed to non-trivial GQTs in higher dimensions, which admit modifications of Schwarzschild as solutions, but not Schwarzschild itself. 


\section{A ``mysteriously'' simple sextic density}\label{Ommm6}
We have seen that all GQT densities as well as all densities trivially satisfying the holographic c-theorem  emanate from a single sextic density, $\Omega_{(6)}$, defined in \req{Omeg6}. The reason for such occurrence can be understood as follows. As we saw earlier, when evaluated on-shell on \req{eq:RGmetric} and \req{fmetric} respectively, the densities $R,\mathcal{S}_2,\mathcal{S}_3$ read\footnote{As a matter of fact, these two ans\"atze have been previously considered simultaneously before in the four-dimensional case \cite{Arciniega:2018fxj,Arciniega:2018tnn,Cisterna:2018tgx} in a cosmological context. The reason is that the condition for demanding a simple holographic c-theorem can be alternatively understood as the condition that the equations of motion for the scale factor in a standard Friedmann-Lema\^itre-Robertson-Walker ansatz  are second order \cite{Sinha:2010pm}.   }
\begin{align}
  \left. R\right|_a      &=   -\frac{2(a'^2 +2 a a'') }{ a^2} \, , \quad
  \left.  \mathcal{S}_2\right|_a  
    = \frac{2(a'^2 - a a'')^2}{3 a^4} \, , \quad
  \left.  \mathcal{S}_3\right|_a  
    = \frac{2(a'^2 - a a'')^3}{9 a^6}\, .\\
\left. R\right|_f &=-\frac{1}{r}\left(r f''+2f'\right) \, , \quad  \left. \mathcal{S}_2\right|_f=\frac{1}{6r^2}\left(r f''- f'\right)^2 \, , \quad  \left.  \mathcal{S}_3\right|_f=\frac{1}{36r^3}\left(r f''- f'\right)^3 \, .
\end{align}
Observe that $\mathcal{S}_2$ and $\mathcal{S}_3$ have in both cases the same functional dependence on $f(r)$ and $a(\rho)$, respectively, up to a power, whereas $R$ has a different dependence from the other two densities in both cases. Now, for $n\leq 5$, there is no way to construct linear combinations of the various order-$n$ densities such that the resulting expression identically vanishes. This is not the case for $n=6$. In that case, $\mathcal{S}_2^3$ and $\mathcal{S}_3^2$ have exactly the same functional dependence on $f(r)$ and $a(\rho)$ respectively, and a particular linear combination of them can be found such that it identically vanishes. This combination is precisely $\Omega_{(6)}$ in both cases,
\begin{equation}
\Omega_{(6)} = 6 \mathcal{S}_3 ^2-\mathcal{S}_2 ^3\, , \quad \left. \Omega_{(6)} \right|_{a}=\left. \Omega_{(6)} \right|_{f}=0\, .
\end{equation}
It is obvious that any density multiplied by $\Omega_{(6)}$ will similarly vanish for these two ans\"atze. One could wonder what happens for other values of $n$ such as $n=12,18,\dots$, for which there is again a match in the functional dependence of the seed densities to the corresponding powers. It is however easy to see that in those cases the combinations which vanish are precisely the ones given by powers of $\Omega_{(6)}$.


It is natural to wonder whether $\Omega_{(6)}$ may actually vanish identically for general metrics. This is however not the case. For instance, for a general static black hole ansatz 
\begin{equation}
\diff s^2=-N^2(r)f(r)\diff t^2 +\frac{ \diff r^2}{f(r)}+r^2 \diff \phi^2\, ,
\end{equation}
one finds that  $\Omega_{(6)}$ is a complicated function of $f(r)$ and $N(r)$. 

As it turns out, the particular linear combination $6\mathcal{S}_3 ^2-\mathcal{S}_2^3$ appearing in $\Omega_{(6)}$ is in fact connected to the Segre classification of three-dimensional spacetimes \cite{Hall,Sousa:2007ax}.  This consist in characterizing the different types of metrics according to the eigenvalues of the traceless Ricci tensor $\tilde R_{ab}$. There exist three large sets of metrics which are precisely characterized by the relative values of  $6\mathcal{S}_3 ^2$ and $\mathcal{S}_2^3$, namely \cite{Chow:2009km,Gurses:2011fv,Chow:2009vt}: 
\begin{align}
\rm{Group\,\, 1:} \quad &6\mathcal{S}_3 ^2=\mathcal{S}_2^3=0\, , \quad &[\text{Type-O, Type-N, Type-III}] \\ \label{group2}
\rm{Group\,\, 2:} \quad &6\mathcal{S}_3 ^2=\mathcal{S}_2^3 \neq 0\, , \quad &[\text{Type-D}_s \text{, Type-D}_t \text{, Type-II}] \\
\rm{Group\,\,  3:} \quad &6\mathcal{S}_3 ^2 \neq \mathcal{S}_2^3\, , \quad &[\text{Type-I}_{\mathbb{R}}, \text{Type-I}_{\mathbb{C}}]
\end{align}
The first group, which in the ---perhaps more familiar--- Petrov notation includes Type-O, Type-N and Type-III spacetimes corresponds to spacetimes such that both $\mathcal{S}_3$ and $\mathcal{S}_2$ vanish. The second group, which includes Type-D and Type-II spacetimes is the one corresponding to metrics which have non-vanishing $6\mathcal{S}_3^2$ and $\mathcal{S}_2^3$ but such that they are equal to each other, \ie such that $\Omega_{(6)}=0$. Finally, spacetimes of Type-I  have a non-vanishing $\Omega_{(6)}$. 

Metrics of the Group 1 have traceless Ricci tensors which can be written as
\begin{align}
&\tilde R_{ab}=0 \quad &[\text{Type-O}]\, , \\ 
&\tilde R_{ab}= s \lambda_a \lambda_b\, , \quad &[\text{Type-N}]\, ,\\
&\tilde R_{ab}= 2 s \xi_{(a}\lambda_{b)}\, , \quad & [\text{Type-III}]\, ,
 \end{align}
where 
\begin{equation}
\quad g^{ab}\lambda_a\lambda_b=0\, , \quad  g^{ab}\xi_a\xi_b=1\, , \quad  g^{ab}\lambda_a\xi_b=0\, , \quad s=\pm 1 \,.
\end{equation}
On the other hand, for metrics of the Group 2 we have 
\begin{align}\label{typeD}
&\tilde R_{ab}=p(x^a) \left[g_{ab}-\frac{3}{\sigma} \xi_a \xi_b\right] \quad &[\text{Type-D}_{s,t}]\, , \\ 
&\tilde R_{ab}=p(x^a) \left[g_{ab}-\frac{3}{\sigma} \xi_a \xi_b\right] + s \lambda_a\lambda_b \quad &[\text{Type-II}]\, , 
 \end{align}
where $p(x^a)$ are scalar functions and
\begin{equation}
\quad g^{ab}\lambda_a\lambda_b=0\, , \quad  g^{ab}\xi_a\xi_b=\sigma=\pm 1\, , \quad  g^{ab}\lambda_a\xi_b=0\, , \quad s=\pm 1 \,.
\end{equation}
Finally, metrics of the Group 3 satisfy
\begin{align}
&\tilde R_{ab}=p(x^a) \left[g_{ab}-3 \xi_a \xi_b\right] -q(x^a) [\lambda_a\lambda_b+\nu_a\nu_b]\quad &[\text{Type-I}_{\mathbb{R}}]\, , \\ 
&\tilde R_{ab}=p(x^a) \left[g_{ab}-3 \xi_a \xi_b\right] -q(x^a) [\lambda_a\lambda_b-\nu_a\nu_b]\quad &[\text{Type-I}_{\mathbb{C}}]\, ,
\end{align}
where $p(x^a)$ and $q(x^a)$ are scalar functions (such that $q\neq \pm 3 p$ for Type-I$_{\mathbb{R}}$) and where
\begin{equation}
\quad g^{ab}\lambda_a\lambda_b=0\, , \quad  g^{ab}\xi_a\xi_b=1\, , \quad  g^{ab}\nu_a\nu_b=0\, ,\quad  g^{ab}\lambda_a\xi_b=g^{ab}\lambda_a\nu_b=0\, , \quad g^{ab}\lambda_a\nu_b=-1\, .
\end{equation}

For the single-function black hole metric and the holographic c-theorem metric, one finds that the traceless Ricci tensor satisfies \req{typeD} with
\begin{align}
p(r)=\frac{f'(r)-r f''(r)}{6r}\, , \quad \xi_a =r \delta_{a\phi} \, ,\quad \text{and} \quad
p(\rho)=\frac{a''(\rho) a(\rho)-a'(\rho)^2}{3a(\rho)^2}\, , \quad \xi_a = \delta_{ar} \, ,
\end{align}
respectively. Hence, both spacetimes are of Type-D$_s$ and from \req{group2} it follows that $\Omega_{(6)}=0$ in both cases.

From this we learn that the appearance of $\Omega_{(6)}$ as a distinguished density was to be expected for the classes of metrics considered here, and that a similar phenomenon is likely to occur for all metrics of the Types D and II.\footnote{For various papers classifying and obtaining explicit solutions of the Groups 1 and 2 for three-dimensional higher-curvature gravities, see \cite{Chow:2009km,Gurses:2011fv,Chow:2009vt,Ahmedov:2010uk,Ahmedov:2011yd,Ahmedov:2012di,Alkac:2016xlr,Alkac:2017rxr}.} Still, the fact that all densities satisfying the holographic c-theorem in a trivial fashion and that all GQTs are proportional to this density was far from obvious in advance.

We believe it would be interesting to further study the properties of $\Omega_{(6)}$, understood as a higher-curvature density. 
Its equations of motion can be easily computed using expression \eqref{eq:EOM3}, and read
\begin{align}
\frac{\mathcal{E}_{ab}^{\Omega_{(6)}}}{3}=&-\frac{1}{6}g_{ab}\left(6\mathcal{S}_3^2-\mathcal{S}_2^3\right)-2\mathcal{S}_2^2\tilde{R}_a^c\tilde{R}_{bc}+12\mathcal{S}_3\tilde{R}_a^c\tilde{R}_{cd}S^d_b-4\left(g_{ab}\Box-\nabla_a\nabla_b+\tilde{R}_{ab}+\frac{1}{3}g_{ab}R\right)\mathcal{S}_2\mathcal{S}_3\notag\\
&-g_{ab}\nabla_c\nabla_d\left(\mathcal{S}_2^2\tilde{R}^{cd}-6\mathcal{S}_3\tilde{R}^{cf}\tilde{R}_f^d\right)-\left(\Box+\frac{2}{3}R\right)\left(\mathcal{S}_2^2\tilde{R}_{ab}-6\mathcal{S}_3\tilde{R}_a^c\tilde{R}_{bc}\right)\notag\\&
+2\nabla_c\nabla_{(a}\left(\tilde{R}_{b)}^c\mathcal{S}_2^2-6\tilde{R}_{b)}^d\tilde{R}_d^c\mathcal{S}_3\right)\, .
\end{align}
These are identically satisfied by all metrics belonging to the Groups 1 and 2, but not for Type-I metrics. It would be then within such set that non-trivial solutions would arise.

\section{Final comments}
In this paper we have presented several new results involving general higher-curvature gravities in three dimensions. A summary of our findings can be found in Section \ref{summary}. Let us close with a couple of possible directions which would be in the spirit of the results presented here.

In \cite{Alkac:2020zhg} it was shown that removing the terms involving Weyl tensors of the $D$-dimensional Lovelock densities and taking the $D\rightarrow 3$ limit in the remainder Ricci parts, one is left with three-dimensional densities which precisely match the ones satisfying an holographic c-theorem \cite{Sinha:2010ai,Paulos:2010ke}. The procedure is applied to the $n=2,3,4$ densities, for which no ``trivial'' densities exist. It would be interesting to explore which particular combination of non-trivial densities is selected by this procedure for higher-order densities and what their relation is with the densities $\mathcal{C}_{(n)}$ and $\mathcal{B}_{(n)}$ identified here.

In this paper we have shown that no non-trivial GQT gravities exist in three dimensions. The situation changes when matter fields are included in the game. Following higher-dimensional inspiration \cite{Cano:2020qhy}, in \cite{Bueno:2021krl} we found a family of non-trivial theories linear in the Ricci tensor coupled to a scalar field which becomes a total derivative when evaluated in \req{fmetric} with a magnetic ansatz for the scalar. These ``electromagnetic quasi-topological gravities'' possess solutions which are continuous extensions of the BTZ black hole ---some of which describe regular black holes without any fine-tuning of parameters. 
 In this context, we expect that more theories of the ``electromagnetic generalized quasi-topological''  class should exist when terms involving more Ricci curvatures are considered.
 
Another venue involves the problem of finding theories with reduced-order traced equations. As we mentioned in the introduction, this was explored in \cite{Oliva:2010zd}, where it was shown that NMG is the only quadratic and/or cubic density which has traced equations of second order. Within the same group of densities, it was shown that $C_{abc}C^{abc}$, where $C_{abc}$ is the Cotton tensor, is the only one which has third-order traced equations. The condition is essentially related to the vanishing of the terms involving two explicit covariant derivatives in the field equations (\ref{eomsss}). For theories involving no explicit covariant derivatives in the action  ---as in \req{action3} --- the trace of the equations (\ref{eq:EOM2}) reads
\begin{equation}
-\frac{1}{2}R-\frac{3}{L^2}-\frac{3}{2}\mathcal{F} + R \mathcal{F}_R +2\mathcal{R}_2 \mathcal{F}_{\mathcal{R}_2}+3 \mathcal{R}_3 \mathcal{F}_{\mathcal{R}_3} +\nabla_a\nabla_b Y^{a b}=0,
\end{equation}
where 
\begin{equation}
Y^{a b}\equiv R^{ab}\mathcal{F}_{\mathcal{R}_2}+\frac{3}{2}R^{ac}R_c ^b \mathcal{F}_{\mathcal{R}_3}+g^{ab}\left(2 \mathcal{F}_R+ R \mathcal{F}_{\mathcal{R}_2}+\frac{3}{2}\mathcal{R}_2 \mathcal{F}_{\mathcal{R}_3}\right).
\end{equation}
Hence, theories with traced equations of second order would be those for which the rank-two symmetric tensor $Y^{ab}$ is conserved. As it turns out, for $n=2$ the only theory of this kind is NMG, precisely because $Y^{ab}\propto G^{ab}$, with $G^{ab}$ the Einstein tensor. The question of whether there is any higher order gravity other than NMG with second order traced equations of motion is that of whether it is possible to build a conserved tensor $Y^{ab}$ at $n\geq 3$. A natural candidate would be the tensor appearing in the equations of motion of the most general theory of order $n-1$, which is automatically conserved. However, the analysis of  \cite{Oliva:2010zd} shows that the only possibility for $n=3$ would be the $C_{abc}C^{abc}$ density, which involves explicit covariant derivatives in the action and is therefore excluded from the analysis. Since $Y^{ab}$ itself does not involve explicit covariant derivatives, the same question in $n=4$  would necessarily require the existence of densities of order $n= 3$ whose equations of motion are free from explicit covariant derivatives. The analysis of   \cite{Oliva:2010zd} disproves this possibility and therefore the only chance would be that some other divergence-free rank-two symmetric tensor cubic in curvatures ---not corresponding to the equations of motion of any covariant density--- exists. We believe that such a tensor does not exist  ---which would mean that no other theories with reduced-order traced equations exist amongst $\mathcal{L}(g^{ab},R_{ab})$ theories--- but we have not found a proof of this fact.


\section*{Acknowledgements}
We thank Bayram Tekin, Julio Oliva and Rodrigo Olea for useful discussion. Some of the computations were performed using \texttt{Mathematica} packages \texttt{xAct} \cite{xPerm:2008} and \texttt{xTras} \cite{Nutma:2013zea}, publicly available on \href{http://xact.es}{\texttt{http://xact.es}}. The work of PAC is supported by a postdoctoral fellowship from the Research Foundation - Flanders (FWO grant 12ZH121N). The work of QL is  supported  by the Spanish Ministry of Universities through FPU grant No.  FPU19/04859.  QL also acknowledges financial support from MICINN grant PID2019-105614GB-C22, AGAUR grant 2017-SGR 754, and State Research Agency of MICINN through the "Unit of Excellence Maria de Maeztu 2020-2023" award to the Institute of Cosmos Sciences (CEX2019-000918-M). The work of JM is funded by the Agencia Nacional de Investigaci\'on y Desarrollo (ANID) Scholarship No. 21190234 and by Pontificia Universidad Cat\'olica de Valpara\'iso. The work of GvdV is supported by CONICET and UNCuyo, Inst. Balseiro.

\bibliographystyle{JHEP}
\bibliography{Gravities}

\end{document}

